\documentclass[aps,prl,twocolumn,superscriptaddress,calc,epsfig,10pt]{revtex4-1}
\usepackage{graphicx}
\usepackage{graphics}
\usepackage{float}
\usepackage{amssymb,amsmath}

\usepackage{wasysym}

\usepackage{color}
\usepackage[normalem]{ulem}

\newcommand{\mitCUAaddress}{MIT-Harvard Center for Ultracold Atoms, Cambridge, Massachusetts 02139, USA}
\newcommand{\mitRLEaddress}{Research Laboratory of Electronics, MIT, Cambridge, Massachusetts 02139, USA}
\newcommand{\mitPhysicsAddress}{Department of Physics, MIT, Cambridge, Massachusetts 02139, USA}
\newcommand{\Harvardaddress}{Department of Physics, Harvard University, Cambridge, Massachusetts 02138, USA}
\newcommand{\SanJoseAddress}{Department of Physics and Astronomy, San Jos\'{e} State University, San Jos\'{e}, California 95192, USA}

\newcommand{\ket}{\right\rangle}

\newcommand{\bra}{\left\langle}

\begin{document}

\title{Spin Transport in a Mott Insulator of Ultracold Fermions}
\author{Matthew A. Nichols}
\affiliation{\mitPhysicsAddress}
\affiliation{\mitCUAaddress}
\affiliation{\mitRLEaddress}

\author{Lawrence W. Cheuk}
\affiliation{\mitCUAaddress}
\affiliation{\Harvardaddress}

\author{Melih Okan}
\affiliation{\mitPhysicsAddress}
\affiliation{\mitCUAaddress}
\affiliation{\mitRLEaddress}

\author{Thomas R. Hartke}
\affiliation{\mitPhysicsAddress}
\affiliation{\mitCUAaddress}
\affiliation{\mitRLEaddress}

\author{Enrique Mendez}
\affiliation{\mitPhysicsAddress}
\affiliation{\mitCUAaddress}
\affiliation{\mitRLEaddress}

\author{T. Senthil}
\affiliation{\mitPhysicsAddress}

\author{Ehsan Khatami}
\affiliation{\SanJoseAddress}

\author{Hao Zhang}
\affiliation{\mitPhysicsAddress}
\affiliation{\mitCUAaddress}
\affiliation{\mitRLEaddress}

\author{Martin W. Zwierlein}
\affiliation{\mitPhysicsAddress}
\affiliation{\mitCUAaddress}
\affiliation{\mitRLEaddress}
\date{\today}

\begin{abstract}
Strongly correlated materials are expected to feature unconventional transport properties, such that charge, spin, and heat conduction are potentially independent probes of the dynamics. In contrast to charge transport, the measurement of spin transport in such materials is highly challenging. We observed spin conduction and diffusion in a system of ultracold fermionic atoms that realizes the half-filled Fermi-Hubbard model. For strong interactions, spin diffusion is driven by super-exchange and doublon-hole-assisted tunneling, and strongly violates the quantum limit of charge diffusion. The technique developed in this work can be extended to finite doping, which can shed light on the complex interplay between spin and charge in the Hubbard model.\\
\\
\end{abstract}

\maketitle

In materials, electrons are the elementary carriers of both spin and charge, and one might thus expect that the properties of spin and charge conduction are always closely related. However, strong electron correlations can lead to the separation of charge and spin degrees of freedom, such as in one-dimensional systems~\cite{Giamarchi2004,Auslaender2005,Heeger1988SSH}. The unusual transport properties of the cuprate high-temperature superconductors in the normal state have been proposed to arise from decoupled spin and charge transport~\cite{Anderson1997,lee06hightc}. The simplest model believed to capture the essential features of the cuprate phase diagram, the Fermi-Hubbard model, features spin-charge separation in one dimension~\cite{Lieb1968}. In two dimensions, relevant for the cuprates, strong correlations render calculations of transport properties highly challenging~\cite{Scalapino1993,Bonca1995,Kopietz1998,Mukerjee2006,Hyungwon2012,Snyder2012,Karrasch2014}. Simultaneous measurements of transport in both the charge and spin sectors would thus be of great relevance. However, in the cuprates, creating and manipulating spin currents is difficult.

Cold-atom quantum simulators can be used to experimentally study the Fermi-Hubbard model in a pristine, isolated environment, with full control of all Hubbard parameters~\cite{Esslinger2010FermiHubbard}. The advent of quantum gas microscopes for fermionic atoms~\cite{Cheuk2015,Haller2015,Parsons2015,Omran2015,Edge2015,Brown2017}, with their single-atom, single-lattice site resolution, has enabled precision measurements of the equation of state~\cite{Cocchi2015,hofrichter2015} and of spin and charge correlations~\cite{CheukSpinCharge2016,Boll2016,Parsons2016} of the two-dimensional (2D) Fermi-Hubbard model. These microscopes are poised for the study of transport, as already demonstrated with bosonic atoms~\cite{Cheneau2012Lightcone,FukuharaImpurity2013,Fukuhara2013,Hild2014,Preiss2015,Choi2016}. Previous measurements of fermionic charge transport were performed without the aid of single-atom resolution~\cite{stro07transport,Schneider2012,WXu2016}. However, it has proven difficult to directly connect the observed dynamics of lattice systems to the transport coefficients of the underlying Hamiltonian. Recently, the optical charge conductivity of a dilute, harmonically trapped 3D Fermi-Hubbard system has been measured~\cite{RAnderson2017}, as well as the charge conductance through a mesoscopic lattice in a wire geometry~\cite{Lebrat2017}.

Here, we explored spin transport in the repulsive 2D Fermi-Hubbard model using ultracold fermionic $^{40}$K atoms on a square lattice confined by a uniform box potential. A natural region in the Hubbard phase diagram where spin and charge transport could differ is near the Mott insulator at half-filling, where charge transport is strongly suppressed, whereas spin transport can occur via super-exchange. Previous experiments have studied spin transport in strongly interacting Fermi gases without a lattice, both in three dimensions~\cite{somm11spin,somm11polarontransport,bard14,valtolina2017} and in two dimensions~\cite{koschorreck13,Luciuk2017}. In those studies, spin diffusion was observed to attain the quantum limit of $\sim \hbar/m$, where $\hbar$ is Planck's constant $h$ divided by $2\pi$ and $m$ is the particle mass. Here, we measure both the spin diffusion coefficient $D_{S}$ and the spin conductivity $\sigma_{S}$. These transport coefficients dictate the response of the system to a spin-dependent force and are related through the Einstein relation, $\sigma_{S}=D_{S}\chi$, where $\chi$ is the uniform spin susceptibility, which can be measured independently.

\begin{figure*}[t]
\centering
\includegraphics[scale=1.0]{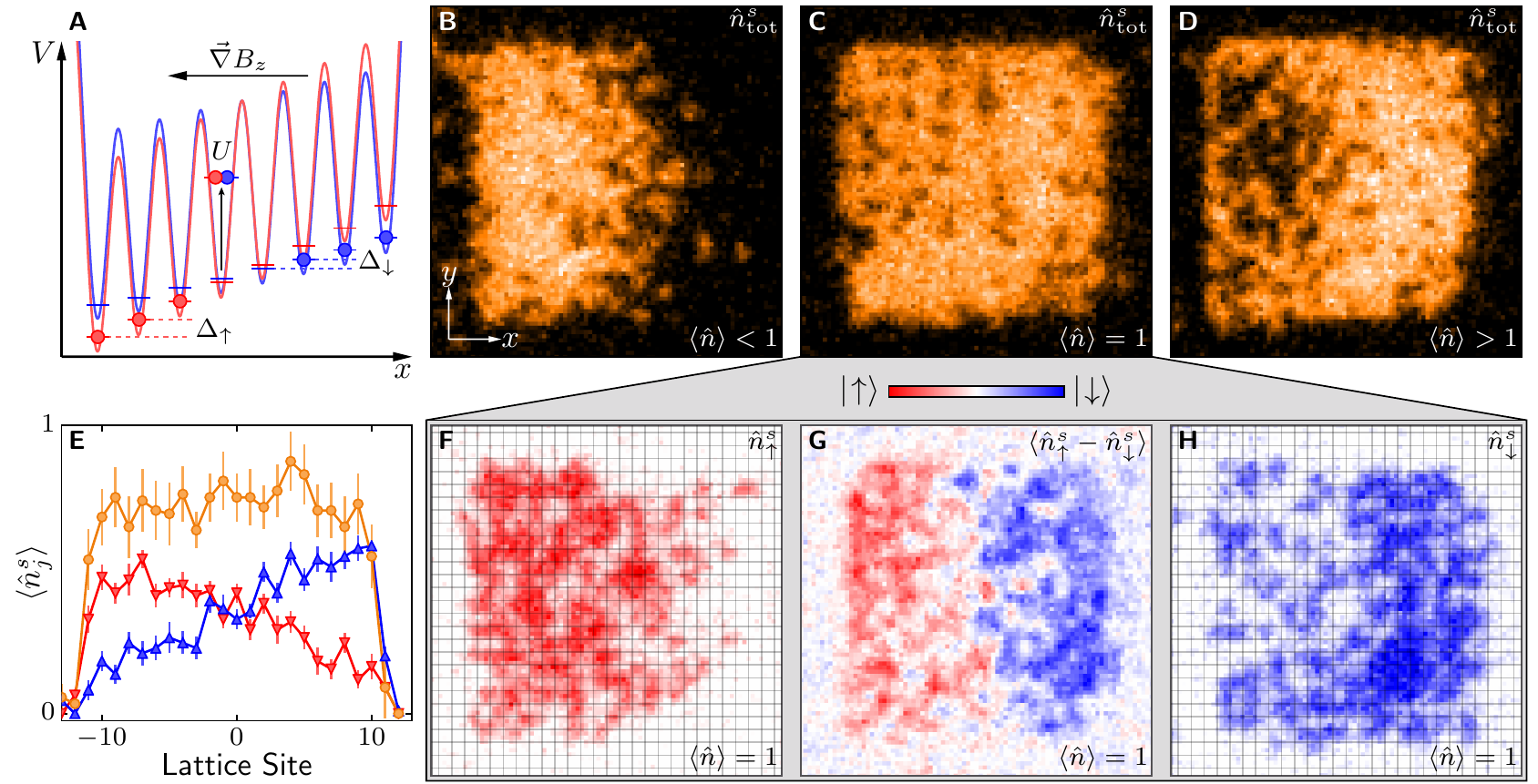}
\caption{\textbf{Creating spin textures in a homogeneous Fermi-Hubbard system.} \textbf{(A)} A diagram of the optical potentials used to confine the atoms, and the tilted lattice potential experienced by the two spin states $\left|\uparrow\ket$ (red) and $\left|\downarrow\ket$ (blue) in the presence of a magnetic field gradient. \textbf{(B-D)} Raw fluorescence images of the parity-projected total density $\hat{n}^{s}_{{\rm tot}}$ for total densities $\bra \hat{n}\ket<1$, $\bra \hat{n}\ket=1$, and $\bra \hat{n}\ket>1$, respectively, which have been prepared adiabatically in the presence of the magnetic gradient; $t/U=0.114(7)$, $0.067(4)$, and $0.114(7)$ for (B), (C), and (D), respectively. \textbf{(E)} The average singles densities, $\bra \hat{n}^{s}_{{\rm tot},j}\ket$ (gold), $\bra \hat{n}^{s}_{\uparrow,j}\ket$ (red), and $\bra \hat{n}^{s}_{\downarrow,j}\ket$ (blue) over four independent realizations at $t/U=0.026(2)$, averaged along the $y$-direction from the reconstructed detected site occupations. Error bars represent $1\sigma$ statistical uncertainty. The average singles densities shown have not been corrected for finite detection fidelity. \textbf{(F)} A single raw image of $\hat{n}^{s}_{\uparrow}$ at $t/U=0.067(4)$. \textbf{(G)} Fluorescence of $\left|\uparrow\ket$ minus fluorescence of $\left|\downarrow\ket$ averaged over six images for the same configuration as (F). \textbf{(H)} A single image of $\hat{n}^{s}_{\downarrow}$ for the same configuration as (F).}
\label{fig1}
\end{figure*}

The 2D Fermi-Hubbard model is realized by evaporatively cooling $^{40}$K atoms to quantum degeneracy and preparing them in an equal mixture of the hyperfine states $\left|\uparrow\ket\,\equiv\,\left|F=9/2, m_F=-3/2\ket$ and $\left|\downarrow\ket\,\equiv\,\left|F=9/2, m_F=1/2\ket$ in a single layer of a highly oblate optical dipole trap~\cite{CheukMott2016}. A sample with uniform filling is produced by projecting a repulsive optical potential through the microscope objective (Fig.~1A), which isolates a uniform $22\times22$ site region of the system~\cite{supplementalmaterial}. The sample is subsequently prepared adiabatically in a square optical lattice, where it is described by the single-band Hubbard Hamiltonian
\begin{eqnarray}
\hat{H} = -t \sum_{\left<i,j\right>,\sigma} \left(\hat{c}_{\sigma,i}^\dagger \hat{c}_{\sigma,j}+h.c.\right) + U \sum_i \hat{n}_{\uparrow,i}\hat{n}_{\downarrow,i}\nonumber \\
- \mu_{\uparrow} \sum_i \hat{n}_{\uparrow,i}-\mu_{\downarrow} \sum_i \hat{n}_{\downarrow,i}\nonumber \\
+ \Delta_{\uparrow} \sum_i i_x\hat{n}_{\uparrow,i} + \Delta_{\downarrow} \sum_i i_x\hat{n}_{\downarrow,i}.
\end{eqnarray}
Here, $t$ and $U$ denote the nearest-neighbor tunneling amplitude and on-site interaction energy, respectively; $\left<i,j\right>$ represents nearest-neighbor sites $i$ and $j$; $\mu_{\uparrow}$ ($\mu_{\downarrow}$) is the chemical potential of atoms in state $\left|\uparrow\ket$ ($\left|\downarrow\ket$); $i_x$ represents the $x$-coordinate of lattice site $i$; and $\Delta_{\uparrow}$ ($\Delta_{\downarrow}$) represents a possible spin-dependent tilt of the potential along the $x$-direction for state $\left|\uparrow\ket$ ($\left|\downarrow\ket$). The operators $\hat{c}_{\sigma,i}^\dagger \left(\hat{c}_{\sigma,i}\right)$ are the fermion creation (annihilation) operators for spin $\sigma=\,\,\uparrow,\downarrow$ on lattice site $i$, and $\hat{n}_{\sigma,i}=\hat{c}_{\sigma,i}^\dagger\hat{c}_{\sigma,i}$ is the number operator on site $i$. To measure the spin transport coefficients $\sigma_{S}$ and $D_{S}$ at half-filling, we apply a spin-dependent force derived from a magnetic gradient along $-\hat{x}$ (Fig.~1A). The magnetic gradient gives rise to a linear tilt in the potential energy of $\Delta_{\uparrow}/h=41.1(8)\,\rm{Hz/site}$ and $\Delta_{\downarrow}/h=15.4(3)\,\rm{Hz/site}$. This tilt has the same sign for atoms of both spins but differs in magnitude. The Hubbard parameters $t$ and $U$ have typical values given by $t/h\sim100\,\rm{Hz}$ and $U/h\sim1\,\rm{kHz}$, and their ratio is varied using the depth of the optical lattice.

\begin{figure*}[t]
\centering
\includegraphics[scale=1.0]{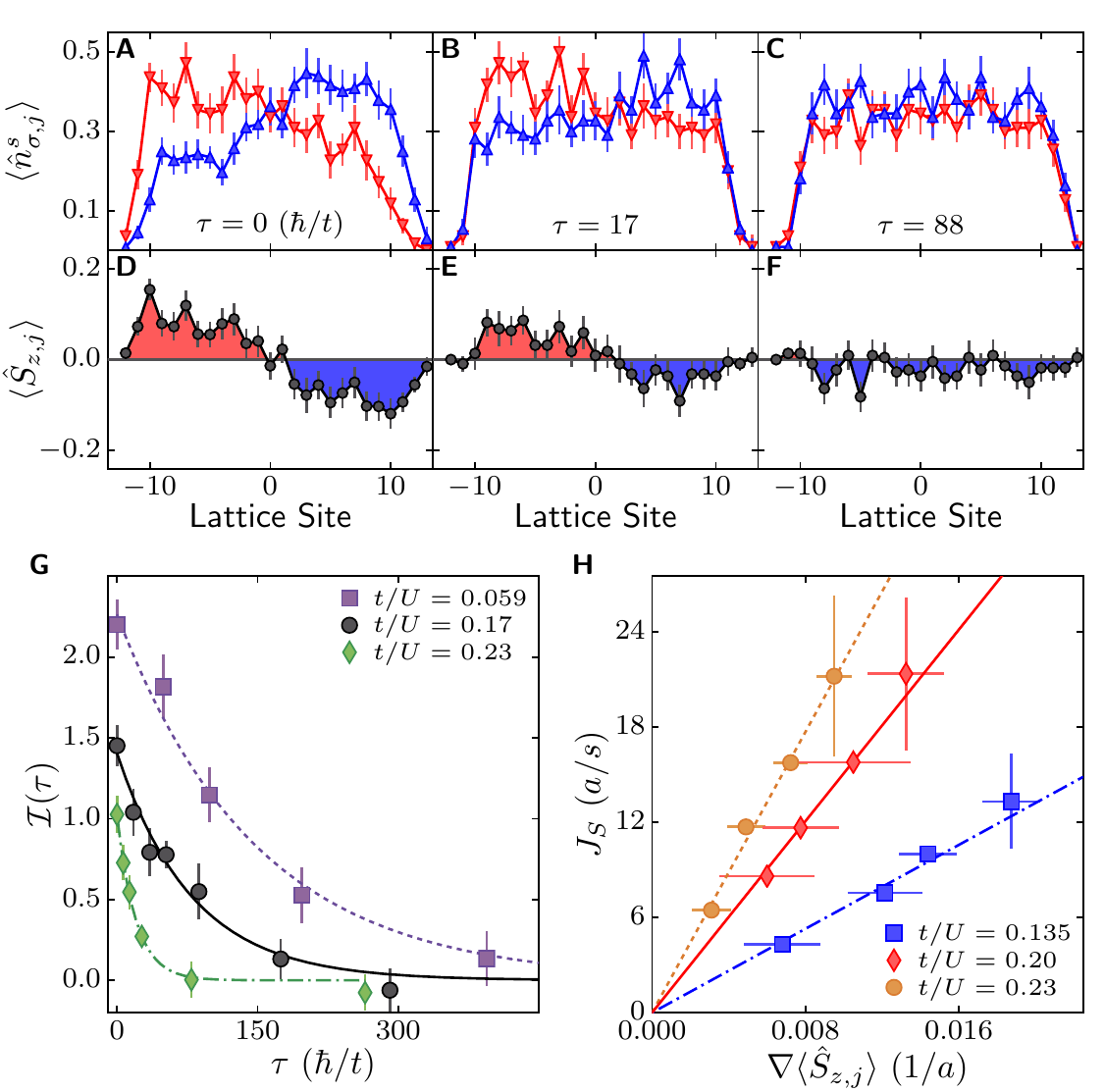}
\caption{\textbf{Observation of spin relaxation after sudden gradient removal.} \textbf{(A-F)} Time evolution of the average singles densities $\bra \hat{n}^{s}_{\uparrow,j}(\tau)\ket$ (red) and $\bra \hat{n}^{s}_{\downarrow,j}(\tau)\ket$ (blue) (upper panels), and of the spin density $\bra \hat{S}_{z,j}(\tau)\ket$ (lower panels), after removing the magnetic field gradient for $t/U=0.23(1)$ at times $\tau/(\hbar/t)=0$, $17$, and $88$. \textbf{(G)} Imbalance $\mathcal{I}(\tau)$ for $t/U=0.059(5)$ (purple squares), $t/U=0.17(1)$ (black circles), and $t/U=0.23(1)$ (green diamonds) and exponential fits to the data. All error bars in (A-G) represent $1\sigma$ statistical uncertainty. \textbf{(H)} Spin current $J_{S}$ at $j=0$ as a function of the spatial gradient in $\bra \hat{S}_{z,j}(\tau)\ket$ at $j=0$ for $t/U=0.135(9)$ (blue squares), $t/U=0.20(1)$ (red diamonds), and $t/U=0.23(1)$ (orange circles) and corresponding linear fits to the data. The error bars along the horizontal axis represent $1\sigma$ statistical uncertainty in the measurement of the spatial gradient in $\bra \hat{S}_{z,j}(\tau)\ket$. Vertical error bars are representative for each curve, derived from the uncertainty in the exponential fit to the imbalance $\mathcal{I}(\tau)$, and are proportional to the magnitude of the spin current. The data in (A-H) have not been corrected for finite detection fidelity.}
\label{fig2}
\end{figure*}

We first measure the spin diffusion coefficient by preparing the sample adiabatically in the presence of the magnetic gradient. The equilibrium density profile can be understood through the local density approximation (LDA). Under LDA, the local chemical potential $\mu_{\sigma,j}$ decreases linearly along the $x$-direction with slope $\Delta_{\sigma}$, for $\sigma=\,\uparrow,\downarrow$. For a weakly interacting system, one expects the densities of both spins to decrease monotonically along $\hat{x}$. This is observed in fluorescence images of samples below and above half-filling, shown in Fig.~1, B and D, respectively. In Fig.~1D, doubly occupied sites appear as holes because of light-assisted collisions during the imaging process~\cite{depu99}, so that the left side of the box region, where the density is highest, appears empty. At half-filling, however, the large charge gap of order $U$ present in the Mott-insulating regime suppresses the formation of double occupancies as long as $\Delta_{\uparrow,\downarrow}\ll\,U$, so that the average density remains homogeneous throughout the sample (Fig.~1, C and E). This directly demonstrates the incompressibility of the Mott-insulating state, which, in an isolated system, suppresses the transport of charge. Spin transport, on the other hand, is not impeded, as spins are free to move.

Indeed, although the total density is insensitive to position, the individual spin densities reveal the effect of the gradient. As shown in Fig.~1E, as well as through images of the individual spin states in Fig.~1, F to H, we observe that $\left|\uparrow\ket$ spins accumulate toward $-\hat{x}$, whereas $\left|\downarrow\ket$ spins accumulate toward $+\hat{x}$~\cite{Weld2009}. The incompressibility of the Mott insulator forces $\left|\downarrow\ket$ spins to occupy the right half of the sample at the expense of an increase in potential energy due to the tilt. The thermodynamic properties, including individual spin densities and double occupancies, of such a tilted fermionic Hubbard system have been studied theoretically using determinant quantum Monte Carlo (DQMC) for weak to intermediate interactions ($0.08<t/U<1$) and gradient strengths comparable to those used in the present work~\cite{Batrouni2017}. Experimentally, we use this separation of the individual spin densities to measure the entropy of the sample; that is, from the equilibrated total spin density profile $\bra \hat{S}_{z,j}\ket=\frac{1}{2} \bra \hat{n}_{\uparrow,j}-\hat{n}_{\downarrow,j}\ket$ in the tilted potential (we retain only the site index along $\hat{x}$), we can obtain the uniform spin susceptibility $\chi=\frac{\partial\bra \hat{S}_{z,j}\ket}{\partial\Delta\mu}$, where $\Delta\mu=\mu_{\uparrow}-\mu_{\downarrow}$, of the unperturbed system in linear response~\cite{supplementalmaterial}. By comparing the measured values of $\chi$ with calculations from the numerical linked-cluster expansion (NLCE) technique~\cite{Rigol2006}, we can determine the entropy per particle $S/k_{B}N$ (where $k_{B}$ is the Boltzmann constant). We find an entropy per particle of $S/k_{B}N=1.1(1)$, a regime where NLCE is expected to converge at half-filling over the range of $t/U$ explored here~\cite{Khatami2011,Khatami2012,CheukSpinCharge2016}.

The equilibrated samples with a spin density gradient provide the starting point for subsequent measurements. Because the initial spin density gradient is small, it acts as a small perturbation to the untilted scenario, ensuring that we are probing properties of the homogeneous system in linear response. After the sample has been prepared at a fixed value of $t/U$, the magnetic gradient is suddenly switched off. Following this quench, the system begins to relax back to equilibrium, where $\bra \hat{S}_{z,j}\ket=0$ everywhere. Figure 2, A to F, shows the decay of the spin density gradient after the quench for $t/U=0.23(1)$. This relaxation implies that a spin current $J_{S}$ must be present. To obtain $J_{S}$ from the measured spin profiles, we define the spin density imbalance, $\mathcal{I}(\tau)$, at time $\tau$ after the quench as
\begin{equation}
  \mathcal{I}(\tau)=\sum_L\bra \hat{S}_{z,j}(\tau)\ket-\sum_R\bra \hat{S}_{z,j}(\tau)\ket
\end{equation}
where $\sum_{L,R}$ denotes summation over the left and right halves of the box. Using the continuity equation for the spin density, one can relate $\mathcal{I}(\tau)$ to the spin current $J_{S}$ at the center of the box $(j=0)$ via $J_{S}(\tau)=-\frac{a}{2}\frac{d}{dt}\mathcal{I}(t)\Bigr|_{\tau}$, where $a$ is the lattice spacing.

\begin{figure*}[t]
\centering
\includegraphics[scale=1.0]{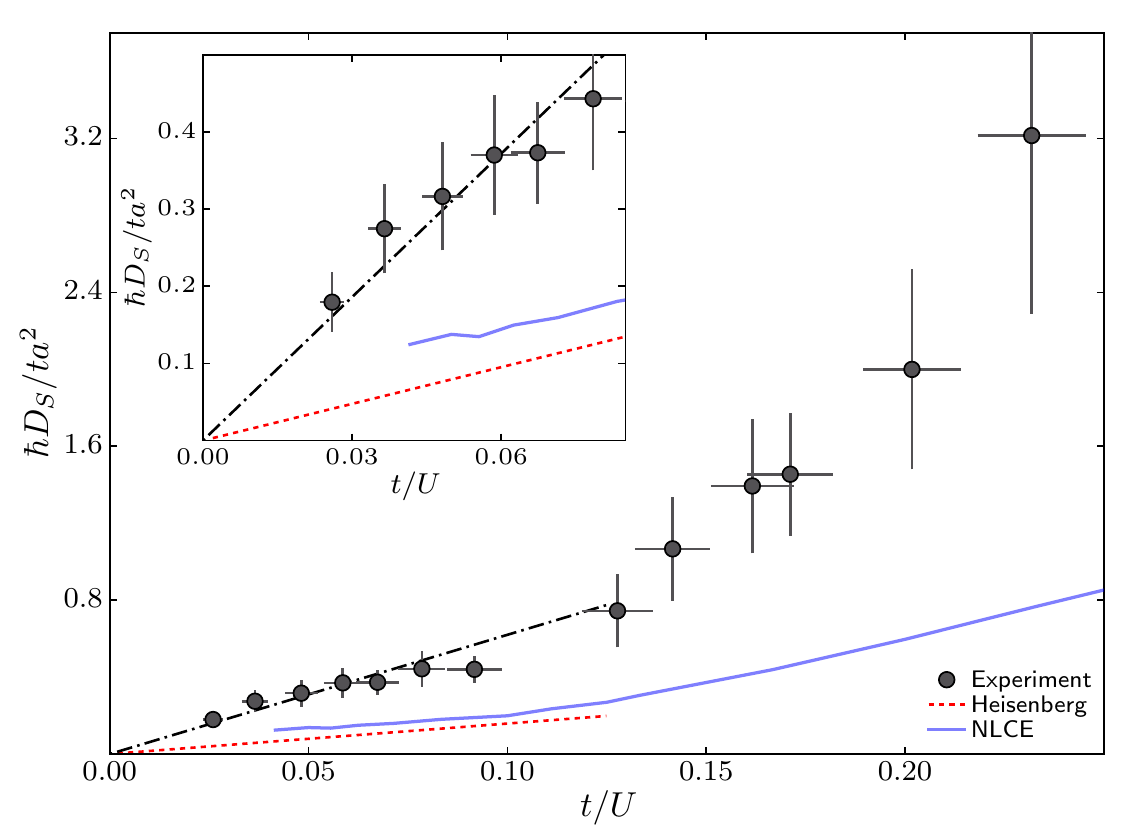}
\caption{\textbf{Spin diffusion coefficient of the half-filled Fermi-Hubbard system versus $t/U$.} The experimentally measured spin diffusion coefficient $\hbar D_S/(ta^{2})\equiv\,D_S/D_0$ at half-filling (black circles) versus the Hubbard parameters $t/U$, and a linear fit to data points with $t/U<0.09$ (black dot-dashed line). The vertical error bars represent the $1\sigma$ statistical error in the measurement; the horizontal error bars represent the $1\sigma$ statistical error in the calibrated value of $t/U$. The blue solid line represents isentropic results for $D_S/D_0$ obtained from NLCE calculations of the real-time spin current-current correlation function for the Hubbard model~\cite{supplementalmaterial}, with an entropy per particle of $1.1 k_B$. With a finite temporal cutoff of $\sim\,$$\hbar/t$ for the real-time correlation functions, the NLCE theory is expected to provide a lower bound to the true diffusivity. For comparison, a prediction for the spin diffusion coefficient of the 2D Heisenberg model at high temperatures, $k_{B}T\gg\,J_{{\rm ex}}$, where $T$ is the temperature, is shown (dashed red line)~\cite{Bennett1965,Sokol1993,Bonca1995}. Inset: A close-up view of the spin diffusion coefficient at half-filling for $t/U<0.09$, where it is expected to scale approximately linearly with $t^2/U$.}
\label{fig3}
\end{figure*}

Figure 2G shows $\mathcal{I}(\tau)$ measured for several values of $t/U$. For all values of $t/U$ explored, $\mathcal{I}(\tau)$ decays to zero. We have verified that the effects of lattice heating during this decay are negligible relative to the experimental uncertainty in the measurement~\cite{supplementalmaterial}. $\mathcal{I}(\tau)$ is then fitted to an exponential curve, and the spin current $J_S$ is obtained through the time derivative of the fit. To connect $J_S$ with the spin transport coefficients, we first examine the dependence of $J_{S}$ on the spin density gradient at the center of the box, $\nabla\bra \hat{S}_{z,j=0}\ket$. By extracting both quantities for a fixed $t/U$ at various times $\tau$, we have access to the dependence of $J_{S}$ on $\nabla\bra \hat{S}_{z,j=0}\ket$ over a large range of values (Fig.~2H). We find that to within experimental error, $J_{S}$ is linearly proportional to $\nabla\bra \hat{S}_{z,j=0}\ket$. This implies that the spin dynamics are diffusive, so that $J_{S}=D_{S}\nabla\bra \hat{S}_{z,j=0}\ket$, where $D_{S}$ is the spin diffusion coefficient. The diffusive nature of the dynamics is also independently probed by a measurement of the power-law dependence of the decay time of $\mathcal{I}(\tau)$, at a fixed value of $t/U$, on the system size $L$~\cite{supplementalmaterial}.

Figure 3 shows the measured spin diffusion coefficient $D_S$ of the half-filled, homogeneous Hubbard model as a function of $t/U$, in units of the quantum scale for mass diffusion $D_0 = \hbar/m$, where $m = \hbar^{2}/t a^2$ is the effective mass in the tight-binding limit. For all data in the strongly interacting regime ($t/U\leq0.125$), the spin diffusion coefficient lies below the scale of quantum-limited mass diffusion $D_0$. In this range, the dependence of $D_S/D_0$ on $t/U$ is linear, implying $D_{S} \propto t^2/U$. From a linear fit constrained to go to zero diffusion at $t/U=0$ (Fig.~3), we obtain $\hbar D_S = 6.2(5)\,a^2 t^2/U$. This $t^2/U$ scaling can be partially understood by considering the Heisenberg limit of the half-filled Fermi-Hubbard model, where spins interact with an exchange coupling $J_{{\rm ex}}=4t^{2}/U$ called the super-exchange energy. Because $J_{{\rm ex}}$ sets the energy scale in this limit, the effective spin mass is given by $m_{S}\sim\,\hbar^{2}/J_{{\rm ex}}a^2\sim\,mU/t$~\cite{Hild2014}. Spin excitations are thus parametrically more massive than $m$. For quantum-limited transport, the spin diffusion coefficient $D_{S}$ is given by $\hbar/m_S$, giving rise to the $t^2/U$ scaling. Although this argument gives the correct scaling, the Heisenberg prediction for the spin diffusion coefficient at temperatures much larger than $J_{{\rm ex}}$ is
\begin{equation}
  \hbar\,D_{S}=4\sqrt{\pi/20}\, a^{2} t^{2}/U \approx 1.6\,a^{2} t^{2}/U
\end{equation}
\cite{Bennett1965,Sokol1993,Bonca1995}, lower than experimentally observed (Fig.~3). This is not surprising, as the Heisenberg model does not capture quantum or thermal doublon-hole fluctuations of the Fermi-Hubbard model, which arise from states with energies greater than $U$~\cite{Kopietz1998}. Doublon-hole fluctuations can increase spin diffusion because spins can move directly from occupied to empty sites, or can trade places with doublons; both processes occur at a rate set by $t$. Because doublon-hole fluctuations are admixed into the wave function of the system with an amplitude proportional to $t/U$ in the strongly interacting regime, the overall scaling of this mechanism is again proportional to $t^{2}/U$. As shown in Fig.~3, for weaker interaction strengths ($t/U>0.125$), the diffusivity $D_S/D_0$ increases faster with $t/U$ than what is given by this initial linear slope.

To gain further insight, we developed a method to calculate the spin conductivity and diffusivity through real-time current-current correlation functions within the NLCE technique~\cite{supplementalmaterial}. This method avoids the ill-posed problem of analytic continuation from imaginary-time, as required in DQMC, and is immune to finite-size effects. These calculations thus give unbiased estimates of transport coefficients in the thermodynamic limit. When comparing the experimental data to the calculations, the only fixed parameter is the entropy per particle, which is independently determined from the measured uniform spin susceptibility. As shown in Fig.~3, the theoretical estimate of the spin diffusivity (blue curve) captures the essential behavior of the experimental data as a function of $t/U$. However, the theoretical calculations systematically underestimate the experimental diffusion coefficient. One possible source of this discrepancy arises from limited access to real-time correlation functions for times longer than $\sim\,$$\hbar/t$. In practice, a cutoff on the order of $\sim \hbar/t$ is used when calculating the direct current (DC) transport coefficients, which can lead to systematic errors. For example, in the Heisenberg limit, one expects real-time correlations to extend out to times $\sim \,\hbar/J_{{\rm ex}}$, which can be much longer than $\hbar/t$. It is therefore notable that even with access to real-time correlations only up to times $\sim\,\hbar/t$, the NLCE estimates agree qualitatively with the experimental data, and also quantitatively to within a factor of $\sim 2$. Although it is difficult to estimate the magnitude of the systematic error, we expect the NLCE estimates to provide a lower bound for $D_{S}$~\cite{supplementalmaterial}.

\begin{figure*}[t]
\centering
\includegraphics[scale=1.0]{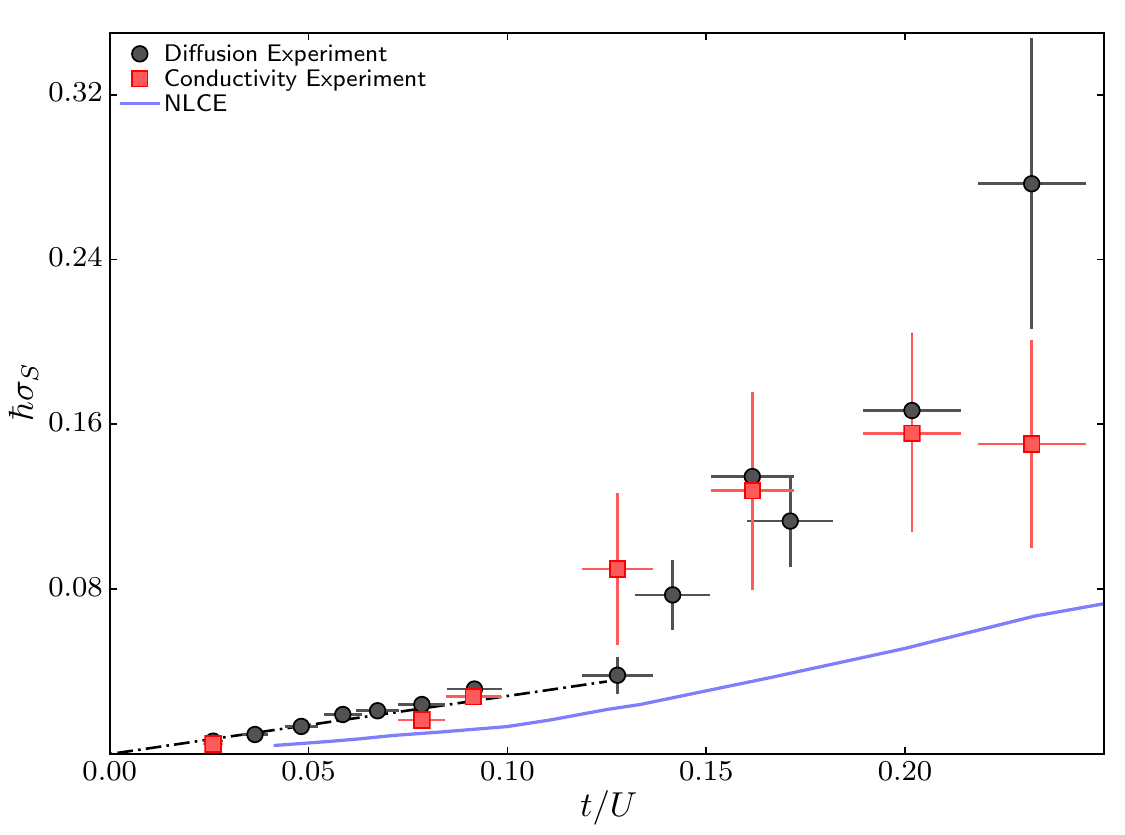}
\caption{\textbf{Spin conductivity of the half-filled Fermi-Hubbard system versus $t/U$.} The measured spin conductivity at half-filling from the initial spin current in an applied magnetic gradient (red squares) and from the measured spin diffusion coefficient using the Einstein relation $\sigma_{S} =D_{S}\chi$ (black circles). A linear fit to data points with $t/U<0.09$ is represented by the black dot-dashed line. The vertical error bars represent the $1\sigma$ statistical uncertainty of the measurements; the horizontal error bars represent the $1\sigma$ statistical error in the calibrated value of $t/U$. The data have been corrected for finite detection fidelity associated with the imaging process of the two spin states~\cite{supplementalmaterial}. The blue solid line is the result obtained for $\hbar\,\sigma_{S}$ at constant entropy using an NLCE calculation of the real-time spin current-current correlation function for the Hubbard model, with an entropy per particle of $1.1 k_B$~\cite{supplementalmaterial}.}
\label{fig4}
\end{figure*}

In addition to the spin diffusion coefficient $D_{S}$, we also independently measure the spin conductivity $\sigma_{S}$. To do this, we first prepare an equilibrated system at half-filling without a potential tilt. We then switch on the tilt suddenly, which induces a spin current in the system. Because $\nabla\bra \hat{S}_{z,j}\ket=0$ at time $\tau=0$, the diffusive contribution to the spin current is negligible initially; in analogy with Ohm's law, which relates a charge current to an applied electric field using the charge conductivity, the initial spin current $J_{S}(\tau=0)$ is directly proportional to the applied spin-dependent force, $-\frac{1}{a}\left(\Delta_{\uparrow}-\Delta_{\downarrow}\right)\hat{x}$, where the spin conductivity $\sigma_{S}$ is the constant of proportionality. Therefore, by measuring the spin current at the center of the box under the known spin-dependent tilt, the spin conductivity can be obtained. The measured spin conductivities at various interaction strengths $t/U$ are shown in Fig.~4. A second way to obtain the conductivity is through the Einstein relation $\sigma_{S}=D_{S}\chi$, where the spin diffusion coefficient $D_S$ and the uniform spin susceptibility $\chi$ are both obtained from the data used in Fig.~3. The values of $\hbar\sigma_{S}$ obtained in this way are also shown in Fig.~4. We find that these two independent methods of measuring the spin conductivity agree with each other to within experimental uncertainty.

We observe that the spin conductivity is linear with $t/U$ in the strongly interacting regime ($t/U\leq0.125$), and find that $\sigma_S = 0.28(2) t/U\hbar$ from a linear fit constrained to yield zero conductivity at $t/U=0$ (Fig.~4). The measured spin conductivities drop far below the Mott-Ioffe-Regel limit for charge in a metal~\cite{Ioffe1960,Mott1972}, $\sigma_0 = n e^2 \tau/m = e^2/\hbar$, derived for a scattering rate $\tau^{-1} = E_F/\hbar$ given by the Fermi energy $E_F$, where the elementary charge of our system is $e=1$. It therefore appears once again as if the effective mass of the carriers of spin is $m_S\sim m U/t$. A breakdown of the Mott-Ioffe-Regel limit is naturally expected in our regime where quasiparticles are ill-defined and Drude-Boltzmann theory does not apply. As $t/U$ increases, the observed spin conductivity grows beyond the initial linear scaling with $t/U$, in analogy with the diffusion coefficient. NLCE predictions for $\hbar\sigma_{S}$ at half-filling (blue curve in Fig.~4) capture the behavior of the spin conductivity with $t/U$ qualitatively, but are systematically lower than the experimental data, for the same potential reasons discussed previously in the context of the diffusion coefficient. Given the substantial challenges associated with calculating the DC limit of the spin conductivity, the experimental data provide a valuable benchmark for future theoretical calculations.

Our study of spin transport can be readily extended in many ways. For example, one can explore the temperature dependence of the spin resistivity, which could display linear behavior reminiscent of charge transport in bad metals. One can also investigate the effect of doping away from half-filling (e.g., at optimal doping), where superconducting fluctuations or a strange metal phase could be present in experimentally attainable conditions. Through simultaneous measurements of both the spin and charge dynamics, such experiments could elucidate the intricate interplay between these two degrees of freedom in the Fermi-Hubbard model.

\begin{acknowledgments}
We thank W. S. Bakr, M. Greiner, and their research groups for fruitful discussions. \textbf{Funding:} Supported by NSF, AFOSR, an AFOSR MURI on Exotic Quantum Phases, ARO, ONR, the David and Lucile Packard Foundation, and Gordon and Betty Moore Foundation grant GBMF5279. E.K. was supported by NSF grant DMR-1609560. The computations were performed in part on the Teal computer cluster of the Department of Physics and Astronomy of San Jos\'{e} State University and in part on the Spartan high-performance computing facility at San Jos\'{e} State University supported by NSF grant OAC-1626645. T.S. was supported by NSF grant DMR-1608505 and partially through a Simons Investigator Award from the Simons Foundation. \textbf{Author contributions:} M.A.N., L.W.C., M.O., T.R.H., E.M., H.Z., and M.W.Z. planned and performed the experiment and analyzed the data. E.K. performed the NLCE simulations. All authors contributed to the interpretation of the data and the preparation of the manuscript. \textbf{Competing interests:} The authors declare no competing financial interests. \textbf{Data and materials availability:} All data shown in this work can be found in an online database~\cite{Nichols2018Dataverse}.
\end{acknowledgments}

\bibliographystyle{apsrev4-1}

\clearpage
\setcounter{equation}{0}
\renewcommand{\theequation}{S\arabic{equation}}
\setcounter{figure}{0}
\renewcommand{\thefigure}{S\arabic{figure}}
\onecolumngrid
\begin{center}
\large{\textbf{Supplementary Materials: \\ Spin Transport in a Mott Insulator of Ultracold Fermions}}
\end{center}
\twocolumngrid

\subsection*{Sample Preparation}

To study spin transport in the 2D Fermi-Hubbard model, we prepare a balanced quantum degenerate mixture of $^{40}$K atoms in the hyperfine states $\left|\uparrow\ket\,\equiv\,\left|F=9/2, m_F=-3/2\ket$ and $\left|\downarrow\ket\,\equiv\,\left|F=9/2, m_F=1/2\ket$ in a single 2D layer of a highly oblate optical dipole trap underneath the high-resolution imaging system. A detailed description of the experimental setup and preparation methods can be found in~(\textit{15,\,43}). Using a magnetic offset field pointing orthogonal to the 2D plane, the magnetic moments of these two states are tuned to a value of $-0.808(5)\,\rm{MHz}/\rm{G}$ and $-0.303(6)\,\rm{MHz}/\rm{G}$, respectively. Subsequently, the atoms are transferred into a single layer of a shallow square optical lattice in the $x$-$y$ plane, with lattice spacing $a=541\,\rm{nm}$, and a depth of $3\,E_R$, where $E_R = \frac{\hbar^2}{2m} \left(\frac{\pi}{a}\right)^2$, and $m$ is the mass of a $^{40}$K atom.

A blue-detuned, repulsive optical potential with a central wavelength of $739\,\rm{nm}$ is then projected onto the atoms through the high-resolution microscope objective using a digital micromirror device (DMD) that is imaged onto the $x$-$y$ plane of the atoms. The optical power in this potential is increased linearly from zero over a $300\,\rm{ms}$ duration. The potential shape is that of a Gaussian profile centered on the atomic cloud, with a square region of size $22$ sites, aligned to the axes of the lattice, removed from the center of the Gaussian profile. This beam isolates a small region of the cloud, providing hard wall boundaries for the atoms within the square, and simultaneously pushes the other atoms in the lattice sufficiently far away, such that they are no longer in contact with the region of interest. The density of atoms in this region can be controlled by varying the number of atoms which are removed during the final stage of evaporative cooling in the optical dipole trap prior to turning on the optical lattice.

\subsection*{Experimental Sequence for Measuring the Diffusion Coefficient}

In order to imprint an equilibrium spin density gradient for the measurement of $D_{S}$ in Fig.~3 of the main text, once the sample is isolated from the rest of the atomic distribution using the projected potential, and the density is set to half-filling, a magnetic field gradient of strength $0.94(2)\,\rm{G/cm}$, which points along the $-\hat{x}-$direction, is turned on adiabatically from zero using a linear ramp of $300\,\rm{ms}$ duration. Due to the fact that the magnetic moments for $\left|\uparrow\ket$ and $\left|\downarrow\ket$ atoms have the same sign, but different magnitudes, this leads to the spin-dependent tilt $\Delta_{\uparrow,\downarrow}$ for the two states discussed in the main text. The magnitude of this tilt is such that $\Delta_{\uparrow,\downarrow}\ll\,U$ for all $U$ explored in this work. We have verified that the experiment operates in the regime of linear response by reducing the gradient by up to a factor of two. This decreased the amplitude of the initial signal in proportion, and yielded the same diffusion coefficient at fixed $t/U$. The spin density profile we imprint is thus a linear perturbation of the homogeneous profile. After the magnetic gradient has reached its final value, the lattices along both $x$ and $y$ axes are increased adiabatically to a depth which yields the desired values of the Hubbard parameters $t/U$. For the purposes of this experiment, the depths of the two axes are kept equal so that the system is 2D. After the end of the lattice ramp, the system is equilibrated in the tilted lattice potential, and a spin density gradient has been established. To initiate the spin dynamics, the magnetic gradient is switched off within $2\,\rm{ms}$, and the spin profile is measured at variable times after the turn-off in order to extract the diffusion coefficient.

\subsection*{Experimental Sequence for Measuring the Conductivity}

An alternate version of this experiment allows us to extract the spin conductivity: after isolating the sample in the box using the projected potential, the lattices along both $x$ and $y$ axes are first increased adiabatically to a depth which yields the desired value of $t/U$. This creates a homogeneous sample where the initial $\mathcal{I}(\tau=0)$  is zero for an equal mixture of the two spin states. Subsequently, the magnetic gradient is turned on from zero within $2\,\rm{ms}$ to its final value of $0.94(2)\,\rm{G/cm}$. This creates a sudden spin-dependent tilt of the lattice potential, which induces spin dynamics by driving the density distributions of the two spin states to separate in the lattice. By measuring the spin profiles at a variable time after the turn-on of the gradient, we obtain the spin conductivity $\sigma_{S}$.

\subsection*{Charge- and Spin-Resolved Detection of Parity Projected Site Occupancies}

To detect the distributions of atoms in the sample at a particular point in time after the dynamics have been initiated, we freeze the atomic distributions by increasing the depth of the square lattice to $\sim100\,E_{R}$ within $2\,\rm{ms}$. Subsequently, an additional $532\,\rm{nm}$ lattice along the $z$-direction, along with the square lattice in the $x-y$ plane, are ramped to $\sim1000\,E_{R}$. We then perform site-resolved fluorescence imaging of the atoms using Raman sideband cooling in order to reconstruct the parity-projected site occupation, which corresponds to a measurement of the singles density $\hat{n}^{s}_{j}=\hat{n}_{\uparrow,j}+\hat{n}_{\downarrow,j}-2\hat{n}_{\uparrow,j}\hat{n}_{\downarrow,j}$~(\textit{15,\,43}). Atom loss during the imaging process results in a finite imaging fidelity of $f=93(2)\%$, which reduces the signal of the average singles density by a factor of $f$.

To perform spin-sensitive imaging for the measurement of the singles density of $\left|\uparrow\ket$ $\left(\left|\downarrow\ket\right)$ atoms, a microwave sweep is used to transfer the atoms in the internal state $\left|9/2, -3/2\ket$ $\left(\left|9/2, 1/2\ket\right)$ to the hyperfine state $\left|7/2, -5/2\ket$ $\left(\left|7/2, 3/2\ket\right)$ after the atomic positions have been frozen at a lattice depth of $\sim100\,E_{R}$. After the microwave sweep, a $5\,\rm{ms}$ pulse of light resonant with the $F=9/2\rightarrow F'=11/2$ transition removes atoms remaining in $F=9/2$ while preserving those in $F=7/2$. The square lattice in the $x-y$ plane, as well as the additional lattice along the $z$-direction with $532\,\rm{nm}$ spacing, are then increased to $\sim1000\,E_{R}$ where we perform Raman imaging to reconstruct the parity-projected occupation of the single spin state. The finite imaging fidelity associated with overall atom loss decreases the signal of the measured singles density for each spin state by a factor of $f$. Additional errors associated with the spin-selective imaging transform the observable operators from $\hat{n}^{s}_{\sigma,j}=\hat{n}_{\sigma,j}-\hat{n}_{\uparrow,j}\hat{n}_{\downarrow,j}$ to $\hat{\tilde{n}}^{s}_{\sigma,j}=(1-\epsilon_{1})\hat{n}_{\sigma,j}+\epsilon_{2}\hat{n}_{-\sigma,j}-(1-\epsilon_{1}+\epsilon_{2})\hat{n}_{\uparrow,j}\hat{n}_{\downarrow,j}$. Here, $\epsilon_{1}$ represents unintended loss of the atoms to be imaged during the $5\,\rm{ms}$ removal pulse, and $\epsilon_{2}$ denotes the error associated with imperfect removal of the other spin state. By measuring the number of atoms remaining after the removal of both spin states, we find that $\epsilon_{2}=0.018(4)$. We estimate that $\epsilon_{1}-\epsilon_{2}=0.11(1)$ by comparing measurements of the average total singles density obtained through normal, spin-independent imaging, with $\bra\hat{\tilde{n}}^{s}_{\uparrow,j}\ket+\bra\hat{\tilde{n}}^{s}_{\downarrow,j}\ket$ obtained using the spin-selective imaging. We thus find that $\epsilon_{1}+\epsilon_{2}=0.15(2)$ for this work. Further details of the spin-sensitive imaging can be found in~(\textit{23}).

\subsection*{Calibration of Hubbard Parameters}

The Hubbard parameter $U$ was calibrated using lattice modulation spectroscopy for a balanced mixture of $\left|\uparrow\ket$ and $\left|\downarrow\ket$ atoms in the absence of the $739\,\rm{nm}$ projected potential. The calibration of the lattice depth was done using lattice modulation spectroscopy to measure the inter-band transition resonance between the ground and first excited bands of the lattice. This was performed on a highly spin polarized sample, where $85\%$ of the atoms were $\left|\downarrow\ket$ and the remaining $15\%$ were $\left|\uparrow\ket$, in order to minimize the possible effects of interactions in the calibration of the depth. The Hubbard parameter $t$ was then extracted based on the measured lattice depth from a tight-binding calculation.

\subsection*{Effects of the Residual Harmonic Confinement}

In the Heisenberg limit of half-filling and $t/U\ll1$, the effective super-exchange coupling between spins on neighboring sites is $J_{ex}=4t^{2}/U$. However, in the presence of a small residual harmonic confinement due to the optical lattice beams, this exchange coupling is modified so that it depends on the site index,
\begin{eqnarray}
  J_{ex,j}=\frac{4t^{2}U}{U^{2}-\left(\frac{1}{2}m\omega^{2}a^{2}\right)^{2}(2j-1)^{2}}.
\end{eqnarray}
Here $\omega$ is the trapping frequency along the $\hat{x}$-direction, in which the spin dynamics occur. For the different $t/U$ values explored in this work, the trapping frequency along the $\hat{x}$-direction varies over a range of $2\pi \times 37(2)\,\rm{Hz}$ to $2\pi \times 64(3)\,\rm{Hz}$. At the maximum trapping frequency, corresponding to the lowest value of $t/U$, $U/h=1077(30)\,\rm{Hz}$. For the sample size chosen for the experiment, the maximum value of $j$ is $11$. With these parameters, the variation of $J_{ex,j}$ over the entire sample is $<1\%$, so that the effects of the harmonic confinement on the dynamics can safely be ignored.

\subsection*{Data Analysis: Extracting the Spin Current and Spin Transport Coefficients}

In each iteration of the experiment, under the constraint of parity-projected imaging, we can access one of the following quantities:
\begin{eqnarray}
 \hat{n}^{s}_{j}= \hat{n}_{\uparrow,j}+\hat{n}_{\downarrow,j}-2\hat{n}_{\uparrow,j}\hat{n}_{\downarrow,j},
\end{eqnarray}
\begin{eqnarray}
 \hat{n}^{s}_{\uparrow,j}= \hat{n}_{\uparrow,j}-\hat{n}_{\uparrow,j}\hat{n}_{\downarrow,j},
\end{eqnarray}
\begin{eqnarray}
 \hat{n}^{s}_{\downarrow,j}= \hat{n}_{\downarrow,j}-\hat{n}_{\uparrow,j}\hat{n}_{\downarrow,j}.
\end{eqnarray}
For this work, we study the averages of these quantities, formed from the average of several (typically $\sim10$) iterations of the experiment for each type of measurement. We thus have access to the average local spin density $\bra \hat{S}_{z,j}\ket=\bra \hat{n}_{\uparrow,j}-\hat{n}_{\downarrow,j}\ket/2=\bra \hat{n}^{s}_{\uparrow,j}-\hat{n}^{s}_{\downarrow,j}\ket/2$ at a given time $\tau$ during the spin dynamics. To obtain this quantity as a function of the site index $j$ along the $x$-direction, we first form the 2D average at each site from multiple iterations of the experiment, and subsequently average along the $y$-direction at each value of $j$. The error bar presented for $\bra \hat{S}_{z,j}\ket$ is the $1\sigma$ statistical uncertainty from these measurements. We then use this to obtain the imbalance $\mathcal{I}(\tau)$ at a particular time $\tau$, which we define as,
\begin{equation}
  \mathcal{I}(\tau)=\sum_{L}\bra \hat{S}_{z,j}(\tau)\ket-\sum_{R}\bra \hat{S}_{z,j}(\tau)\ket.
\end{equation}
The error in $\mathcal{I}(\tau)$ is then the $1\sigma$ statistical uncertainty based on the measurement of $\bra \hat{S}_{z,j}\ket$.

One can then relate $\mathcal{I}(\tau)$ to the spin current $J_{S}(\tau)$ at the center of the box $j=0$ using the continuity equation for $\hat{S}_{z,j}$. That is, from
\begin{equation}
  \frac{d}{d\tau}\hat{S}_{z,\alpha}=\frac{i}{\hbar}[\hat{H},\hat{S}_{z,\alpha}],
\end{equation}
one obtains,
\begin{equation}
  \frac{d}{d\tau}\hat{S}_{z,\alpha}=\frac{it}{\hbar}\sum_{\left<i\right>,\sigma}\sigma(\hat{c}_{\sigma,\alpha}^\dagger \hat{c}_{\sigma,i}-\hat{c}_{\sigma,i}^\dagger \hat{c}_{\sigma,\alpha}),
\end{equation}
where the sum over $\left<i\right>$ runs over the sites which are the nearest-neighbors of site $\alpha$, and $\sigma=(1/2,-1/2)$ for $\left|\uparrow\ket$ and $\left|\downarrow\ket$ respectively. Because we are interested in dynamics only along the $x$-direction, we restrict the remaining discussion to this axis only. Summing over $\alpha$, we find
\begin{eqnarray}
  \frac{d\mathcal{I}}{d\tau}&=&\frac{d}{d\tau}\left(\sum_{\alpha<0}\bra\hat{S}_{z,\alpha}\ket-\sum_{\alpha\geq0}\bra\hat{S}_{z,\alpha}\ket\right) \nonumber \\
&=& \bra\sum_{\alpha<0}\frac{d}{d\tau}\hat{S}_{z,\alpha}-\sum_{\alpha\geq0}\frac{d}{d\tau}\hat{S}_{z,\alpha}\ket.
\end{eqnarray}
Using Eq. (S7) in Eq. (S8), and making use of the fact that our experiment utilizes hard wall boundary conditions so that no particles enter or leave the region of interest, one obtains,
\begin{eqnarray}
  \frac{d\mathcal{I}}{d\tau}&=&\frac{2it}{\hbar}\bra\sum_{\sigma}\sigma(\hat{c}_{\sigma,-1}^\dagger \hat{c}_{\sigma,0}-\hat{c}_{\sigma,0}^\dagger \hat{c}_{\sigma,-1})\ket \nonumber \\
&=& -\frac{2}{a}J_{S},
\end{eqnarray}
where $J_{S}$ is the spin current at the center of the box,
\begin{equation}
  J_{S}=\frac{-ita}{\hbar}\bra\sum_{\sigma}\sigma(\hat{c}_{\sigma,-1}^\dagger \hat{c}_{\sigma,0}-\hat{c}_{\sigma,0}^\dagger \hat{c}_{\sigma,-1})\ket.
\end{equation}
Thus, from the time derivative of the imbalance $\mathcal{I}$ one can obtain the spin current flowing across the center of the box. The additional factor of $a$ in the definition of the spin current is included so that the equation $J_{S}=D_{S}\nabla\bra\hat{S}_{z,j}\ket$ holds, and arises because the spin density $\bra\hat{S}_{z,j}\ket$ is by definition unit-less. Therefore, the spin current, as it is defined, has units of $a$ per unit time.

\begin{figure}[t]
\centering
\includegraphics[scale=1.0]{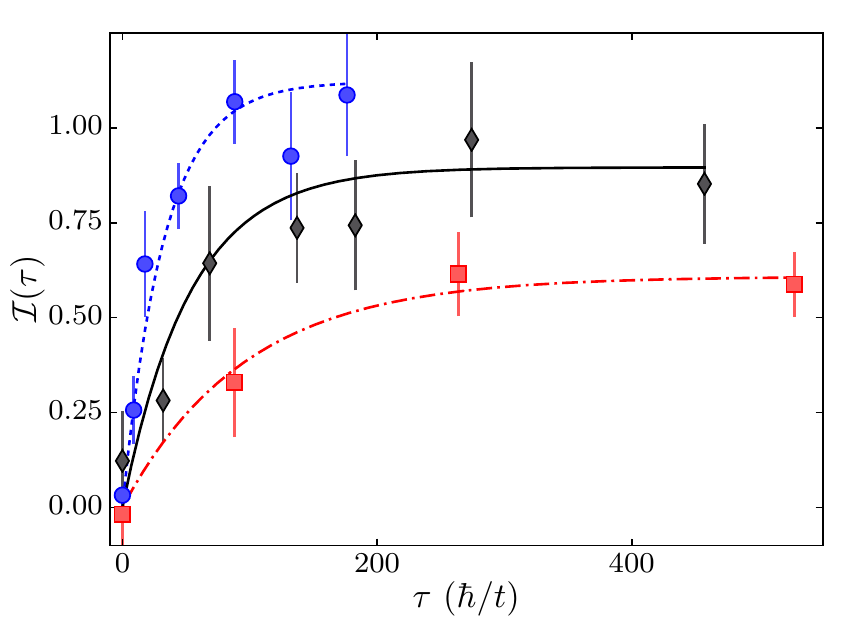}
\caption{{\bf Observation of spin separation following sudden application of gradient.} Time evolution of the imbalance $\mathcal{I}(\tau)$ after abruptly applying a magnetic field gradient to an equilibrium sample for $t/U = 0.026(2)$ (red squares), $t/U = 0.091(5)$ (black diamonds), and $t/U = 0.23(1)$ (blue circles), and exponentially saturating fits to the data.}
\label{figS1}
\end{figure}

To obtain $d\mathcal{I}/d\tau$, we fit the measured imbalance $\mathcal{I}(\tau)$ to an exponential to extract the fit parameters. For the diffusion measurements, where the imbalance decays exponentially as in Fig.~2G of the main text, we fit a function of the form
\begin{equation}
  \mathcal{I}(\tau)=A_{D}e^{-\tau/\tau_{D}},
\end{equation}
to extract the initial amplitude $A_{D}$ and the decay time $\tau_{D}$. The offset of this fit function has been fixed to zero since the real offset can be experimentally controlled well within the measurement error bars. For the conductivity measurements in Fig.~4 of the main text, where the dynamics occur in the presence of the applied gradient, the imbalance begins at zero at $\tau=0$, and then undergoes exponential saturation as the system equilibrates in the tilted potential. This is demonstrated in Fig.~S1 for several values of $t/U$. Because we are interested in the initial slope of these curves in order to extract the spin conductivity, we perform an exponential fit to this data of the form,
\begin{equation}
  \mathcal{I}(\tau)=R_{\sigma}\tau_{\sigma}\left(1-e^{-\tau/\tau_{\sigma}}\right),
\end{equation}
to extract the initial slope $R_{\sigma}$, and the time constant for the saturation $\tau_{\sigma}$.

The transport coefficients $D_{S}$ and $\sigma_{S}$ can then be obtained from the parameters of these fits, as well as Eq. (S9). For the measurement of spin diffusion, the spin current at $\tau=0$, using Eqs. (S9) and (S11), is $J_{S}(\tau=0)=aA_{D}/2\tau_{D}$. Thus, because the spin dynamics occur in the absence of a gradient, so that the spin current has the form $J_{S}=D_{S}\nabla\bra\hat{S}_{z,j}\ket$, where $\nabla\bra\hat{S}_{z,j}\ket$ is the spatial gradient of the spin density profile, the spin diffusion coefficient can be expressed as,
\begin{equation}
  \frac{\hbar\,D_{S}}{ta^{2}}=\frac{\hbar\,A_{D}}{2\tau_{D}t}\frac{1}{\left(\left.\partial\bra\hat{S}_{z,j}\ket/\partial\,j\right)\right|_{j=0,\tau=0}}.
\end{equation}
The initial slope of $\bra\hat{S}_{z,j}\ket$ with respect to lattice index $j$, $\left(\left.\partial\bra\hat{S}_{z,j}\ket/\partial\,j\right)\right|_{j=0,\tau=0}$, is obtained by fitting a line to the measured profile $\bra\hat{S}_{z,j}\ket$ using the $\sim7-10$ sites centered around $j=0$.

We can obtain the spin conductivity using Eqs. (S9) and (S12), as well as the fact that the initial spin current $J_{S}(\tau=0)$ has no contribution from diffusion, since $\nabla\bra\hat{S}_{z,j}\ket=0$ at $\tau=0$, so that it is strictly determined by the applied differential force and the spin conductivity. The spin conductivity can then be expressed as,
\begin{equation}
  \hbar\sigma_{S}=\frac{\hbar\,R_{\sigma}}{2}\frac{1}{\Delta_{\uparrow}-\Delta_{\downarrow}}.
\end{equation}
The experimental values obtained for $\sigma_{S}$ using Eq. (S14) can be compared to the spin conductivity predicted by the Einstein relation, $D_{S}\chi=\sigma_{S}$, using the values of the diffusion coefficient obtained from Eq. (S13) and the uniform spin susceptibility from Eq. (S16), in order to verify that the two versions of the experiment agree. Using the Einstein relation, the spin conductivity can be expressed in terms of quantities obtained from the diffusion measurements,
\begin{equation}
  \hbar\sigma_{S}=\frac{\hbar\,A_{D}}{2\tau_{D}}\frac{1}{\Delta_{\uparrow}-\Delta_{\downarrow}}.
\end{equation}
The errors bars for the experimental data obtained using Eqs. (S13), (S14), and (S15), shown in Fig.~3 and Fig.~4 of the main text, are the $1\sigma$ statistical uncertainties of the measurements. Additionally, the data for the conductivity in Fig.~4 of the main text have been corrected for a finite imaging fidelity of $f=93(2)\%$ associated with atom loss during the imaging process, and a finite spin imaging fidelity arising from the spin detection error $\epsilon_{1}+\epsilon_{2}=0.15(2)$. Whereas $D_{S}$ is independent of both $f$ and $\epsilon_{1}+\epsilon_{2}$ because both $A_{D}$ and $\left(\left.\partial\bra\hat{S}_{z,j}\ket/\partial\,j\right)\right|_{j=0,\tau=0}$ are proportional to $f(1-\epsilon_{1}-\epsilon_{2})$, the experimental conductivity is linearly proportional to $f(1-\epsilon_{1}-\epsilon_{2})$ because both $R_{\sigma}\propto\,f(1-\epsilon_{1}-\epsilon_{2})$ and $A_{D}\propto\,f(1-\epsilon_{1}-\epsilon_{2})$ in Eqs. (S14) and (S15).

\begin{figure}[t]
\centering
\includegraphics[scale=1.0]{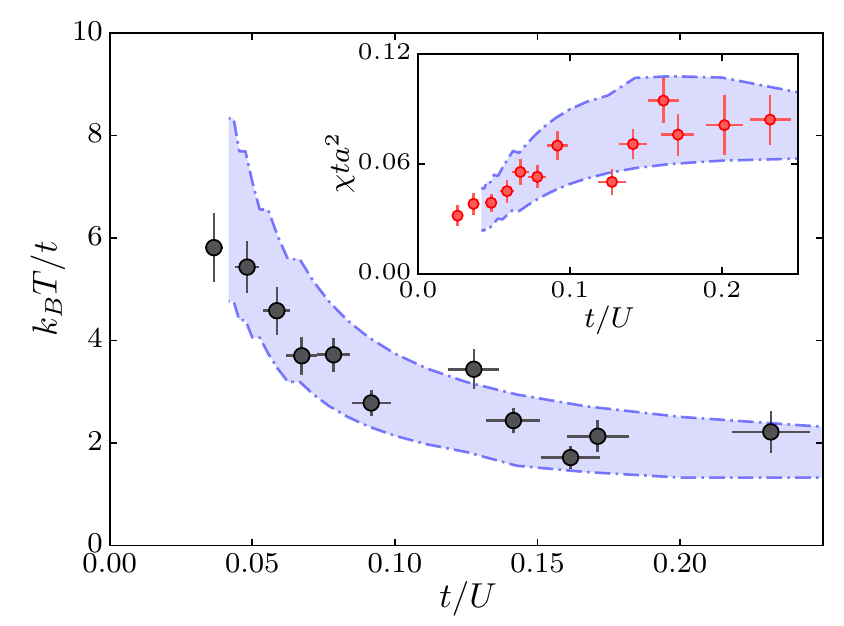}
\caption{{\bf Measured temperature versus $t/U$.} Normalized temperature $k_{B}T/t$ (black points) from the experimentally measured uniform spin susceptibility $\chi$ (inset) versus the Hubbard parameters $t/U$. The shaded blue regions represent NLCE calculations for fixed entropy per particle of $S/k_{B}N = 1.0$ to $S/k_{B}N = 1.2$.}
\label{figS2}
\end{figure}

Throughout this discussion, we have ignored the possibility of thermoelectric effects which could generate thermal gradients during the spin dynamics. This is justified for a spin-balanced system at half-filling, the regime considered for this work, where these effects are minimized. Indeed, the two sources of thermoelectric coupling arising from the spin dynamics are the spin-analog of the thermoelectric susceptibility, $\zeta_{S}=\left.\partial\bra\hat{S}_{z}\ket/\partial\,T\right|_{\mu_{\uparrow},\mu_{\downarrow}}$, and the spin Seebeck coefficient. The spin thermoelectric susceptibility $\zeta_{S}$, a thermodynamic quantity that describes the magnitude of induced spin density gradients originating from gradients in the temperature, is exactly zero for a spin-balanced system at half-filling. This follows from a particle-hole symmetry of the Hubbard model about the point $\mu_{\uparrow}=\mu_{\downarrow}=U/2$, which enforces $\bra\hat{n}_{\uparrow}\ket=\bra\hat{n}_{\downarrow}\ket=0.5$, so that $\bra \hat{S}_{z}\ket=0$, at all temperatures. Similarly, the spin Seebeck coefficient, which describes coupling between the heat and spin transport modes of the system, is exactly zero for a spin-balanced system at half-filling due to spin-rotational invariance at this point. In other words, for a spin-balanced system spin transport is decoupled from both net density and heat transport due to symmetry of the system under $180^{\circ}$ spin-rotations which change the sign of $\hat{S}_{z}$, meaning the spin Seebeck coefficient must vanish~(\textit{11}). Because we operate in the linear regime about the half-filling point for this work, thermoelectric effects due to both $\zeta_{S}$ and the spin Seebeck coefficient vanish to leading order.

\subsection*{Measurement of $\chi$ Versus $t/U$}

Measurements of the initial spin density profile $\bra \hat{S}_{z,j}\ket=\bra \hat{n}^{s}_{\uparrow,j}-\hat{n}^{s}_{\downarrow,j}\ket/2$ for the system equilibrated in the tilted lattice potential prior to the shutoff of the magnetic gradient can be used to extract the uniform spin susceptibility of the unperturbed system at half-filling. This follows from the fact that $\chi=\partial\bra \hat{S}_{z,j}\ket/\partial\Delta\mu$ where $\Delta\mu=\mu_{\uparrow}-\mu_{\downarrow}$, and that $\partial/\partial\Delta\mu$ can be converted to a derivative with respect to lattice index $j$ using the local density approximation. That is, one can obtain a dimensionless expression for the uniform spin susceptibility in terms of experimentally measurable quantities as, 
\begin{eqnarray}
  \chi\,ta^{2}=\frac{\partial\bra \hat{S}_{z,j}\ket}{\partial\,j}\frac{t}{\partial\Delta\mu/\partial\,j}.
\end{eqnarray}
Here, $\partial\Delta\mu/\partial\,j$ is fixed, and is known from the calibrated values of $\Delta_{\uparrow}$ and $\Delta_{\downarrow}$, and $t$ is obtained from the calibration of the Hubbard parameters. Thus, from the slope of $\bra \hat{S}_{z,j}\ket$ for the initial spin profile, we can extract $\chi\,ta^{2}$ as a function of $t/U$ for the unperturbed system at half-filling. This is shown in the inset of Fig.~S2 using the initial spin density profiles of the same data used to obtain $D_{S}$ in Fig.~3 of the main text. The values of $\chi$ obtained in this way can then be used to obtain the spin conductivity $\sigma_{S}$ from the spin diffusion coefficient using $D_{S}\chi=\sigma_{S}$, shown in Fig.~4 of the main text.

The values obtained for $\chi$ can also be used as a thermometer to obtain the temperature and entropy of the sample in the lattice as a function of $t/U$. That is, we can compare our experimentally measured $\chi$ with numerical results from NLCE for the uniform spin susceptibility at half-filling performed on a homogeneous system in equilibrium with the same value of $t/U$. For the temperatures obtained in this work, the NLCE predictions are expected to be exact at half-filling~(\textit{23}). A plot of $k_{B}T/t$ obtained in this way is shown in Fig.~S2 versus $t/U$. To get an accurate comparison of the experimental data with the NLCE predictions, the measured values of $\chi$ in the inset of Fig.~S2 are corrected for the finite detection fidelity $f(1-\epsilon_{1}-\epsilon_{2})$. Because the experimental data shown in Fig.~3 and Fig.~4 of the main text is obtained at what is expected to be constant entropy rather than constant temperature, we can obtain a measure of the entropy in our experiment by comparing the data in Fig.~S2 with that obtained from NLCE data for the temperature $k_{B}T/t$ for a homogeneous system at half-filling as a function of $t/U$ at a fixed entropy per particle. The theoretical predictions are also shown in Fig.~S2. This demonstrates that our experimental data is consistent with an entropy per particle in the range $S/k_{B}N=1.0-1.2$. The system entropy obtained using this method agrees with an estimate of the entropy based on a comparison of the experimentally measured average singles density, $\bra \hat{n}^{s}_{\uparrow}+\hat{n}^{s}_{\downarrow}\ket$, at the center of the box with NLCE results for the average singles density at half-filling performed on a homogeneous system in equilibrium with the same value of $t/U$. This experimental estimate of the entropy per particle of the system, $S/k_{B}N=1.1(1)$, is used to fix the entropy in the independent theoretical estimates for $D_{S}$ and $\sigma_{S}$ shown in Fig.~3 and Fig.~4 of the main text.

\subsection*{Effects of Heating}

To check the effect of heating of the sample during the spin dynamics on our measurement of the spin transport coefficients, we prepare a sample in the presence of the magnetic gradient to create an initial inhomogeneous spin profile, as done for the measurements of the spin diffusion coefficient. After the sample has been prepared, we keep the magnetic gradient on, and hold the atoms in the lattice for a variable time. We then measure the spin profile $\bra \hat{S}_{z,j}\ket$ and the imbalance $\mathcal{I}$ as a function of hold-time in the lattice after preparing the sample. Because of the finite heating rate, the temperature of the sample increases as a function of time, which decreases the uniform spin susceptibility, and therefore leads to a decrease in $\mathcal{I}$. Experimentally we find that $\mathcal{I}$ decreases approximately exponentially in time, so that we may perform an exponential fit to obtain the $1/e$ decay time associated with the heating, $\tau_{heating}$. For the lowest value of $t/U$ used in this work ($t/U=0.026(2)$), where the effects of heating are most significant, we find that $\tau_{heating}=4.8(6)\,\rm{s}$, which is still longer than the evolution time at this point, $\tau_{D}=1.5(4)\,\rm{s}$. Because the spin diffusion coefficient $D_{S}$ is proportional to $1/\tau_{D}$, we estimate this effect leads to a correction of the diffusion coefficient of $<30\%$, which is comparable to the experimental uncertainty in the measured value of this point. Given that the heating time $\tau_{heating}$ is much longer than the evolution time $\tau_{D}$ for all $t/U>0.026$, corrections due to heating are negligible compared to the experimental uncertainty of the measurements for these $t/U$ points. Therefore, none of the data presented in the main text have been corrected for heating effects. The existence of this heating can, however, produce a finite, non-zero offset for $D_{S}$ at very small $t/U$, and therefore sets a limit on the smallest $t/U$ we can study.

\subsection*{Changing the System Size to Verify Diffusive Dynamics}

Additional evidence of diffusive spin dynamics comes from the dependence of the decay time, $\tau_{D}$, on the system size, $L$. That is, for the experimental sequence used to measure the diffusion coefficient $D_{S}$, the decay time of the imbalance $\mathcal{I}$ should obey $\tau_{D}\propto\,L^{2}$ if the dynamics are entirely diffusive. This comes from the combination of the continuity equation for $\bra \hat{S}_{z,j}(\tau)\ket$ with the linear relationship between the spin current and the spatial gradient in $\bra \hat{S}_{z,j}(\tau)\ket$. Sub- or super-diffusive dynamics would imply that $\tau_{D}\propto\,L^{\alpha}$ where $\alpha>2$ or $\alpha<2$ respectively. Because we have control over the system size, $L$, using the DMD to vary the size of the box confining the atoms, we can measure $\tau_{D}$ for different values of $L$ with $t/U$ fixed. For this measurement, both the strength of the magnetic gradient and the initial system preparation are held constant. Fig.~S3 plots the measured values of $\tau_{D}$ extracted from the decay of the imbalance, $\mathcal{I}$, at $t/U=0.079(5)$ as a function of system size, $L$, on a double-logarithmic plot. We fit a power law of the form $\tau_{D}\propto\,L^{\alpha}$ to the data in order to extract the exponent, $\alpha$, of the $L$ dependence of $\tau_{D}$. The fit yields a value for this exponent of $\alpha=2.1(6)$, which is consistent, to within experimental uncertainty, with diffusive spin dynamics, and agrees with the observed linear relationship between the spin current and the gradient in the total spin density shown in Fig.~2H of the main text.

\begin{figure}[t]
\centering
\includegraphics[scale=1.0]{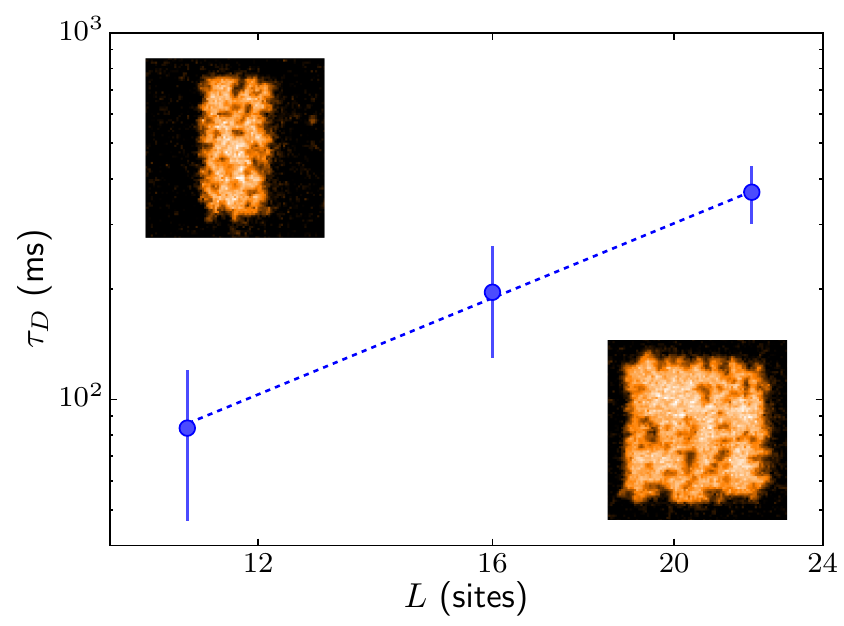}
\caption{{\bf Experimental measurements of the relaxation time $\tau_{D}$ versus the lateral box width.} The dashed line shows a power law fit to the data $\ln(\tau)=\alpha\ln(L)+\beta$ with exponent $\alpha = 2.1(6)$. For the data shown, $t/U = 0.079(5)$. The images are single raw images of the total singles density for the smallest and largest box sizes.}
\label{figS3}
\end{figure}

\subsection*{Theoretical Analysis: Results from the Numerical Linked-Cluster Expansion}

\subsubsection*{Basics of the Method}

We use the NLCE~(\textit{48,\,56}) to obtain estimates for the 
spin conductivity and diffusivity of the Fermi-Hubbard model in two dimensions. NLCE expresses extensive properties of the lattice
model, directly in the thermodynamic limit, as a series in terms of contributions from all finite clusters that can be embedded 
in the lattice. Those contributions are calculated through the inclusion-exclusion principle and within the machine precision using exact diagonalization. 
As the temperature is lowered and the correlations in the system grow larger in order of magnitude than the size of the biggest clusters considered, 
the series will no longer be convergent. However, because of the exact treatment of the system at the 
level of finite clusters, the lowest convergence temperatures typically reach well below the characteristic 
energy scale $t$ and generally decrease at half-filling by increasing $U$ as we approach the atomic limit~(\textit{49,\,50}). 
We carry out our calculations up to the 8th order, where the order indicates the maximum cluster size included 
in the series, and we work in units where $\hbar=a=k_{B}=1$ for the remainder of the discussion.

We have implemented time-dependent (real time as opposed to imaginary time) correlation functions of the Hubbard model at \textit{equilibrium} for the first time in the NLCE to obtain dynamical quantities, such as the spin conductivity, directly in the real frequency domain. Previously, NLCE has been used to study the relaxation dynamics of out-of-equilibrium systems after a quench~(\textit{57,\,58}).

Averaging over the two orientations of every symmetrically distinct cluster in the series, we calculate the uniform 
(${\bf q}=0$) time-dependent spin current-current correlation function, $\left< J_S(\tau)J_S(0)\right>$, along the $x$-direction, where the spin current operator is 
given in Eq. (S10), and use it to obtain the alternating current (AC) spin conductivity through~(\textit{13,\,59})
\begin{equation}
  \textrm{Re } \sigma_S(\omega) = \frac{(1-e^{-\beta \omega})}{\omega} \textrm{Re}\int_0^\infty d\tau e^{i\omega \tau}\left< J_S(\tau)J_S(0)
\right>,\label{eq:sigma3} 
\end{equation}
or equivalently,
\begin{equation}
\textrm{Re }\sigma_S(\omega) = \frac{-2}{\omega} \textrm{ Im} \int_0^\infty d\tau e^{i\omega \tau} \textrm{Im} \left<J_S(\tau)J_S(0)\right>,\label{eq:sigma4}
\end{equation}
where $\beta$ is the inverse temperature.
At temperatures relevant to the experiment, we can safely assume that there is no Drude contribution to the spin conductivity and take
$\sigma_S$ to represents the ``regular" part. We then take the $\omega\to 0$ limit of $\sigma_S(\omega)$ as the direct current 
(DC) spin conductivity.

\begin{figure*}[t]
\centering
\includegraphics[width=6 in]{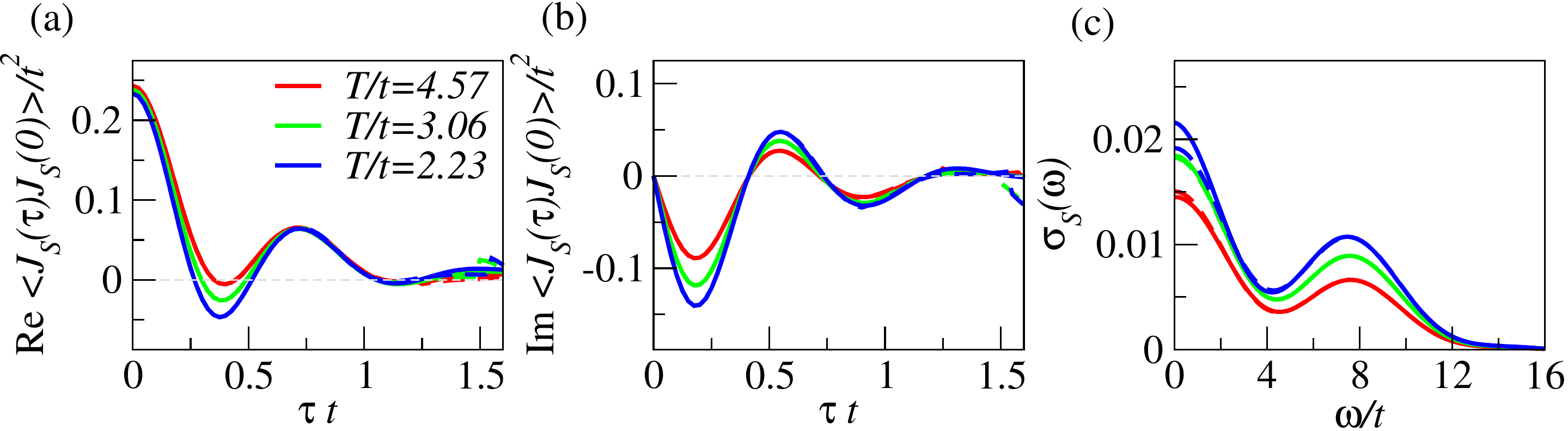}
\caption{{\bf Real-time spin current correlation functions and spin conductivity.} \textbf{(A)} Real part of the time-dependent spin 
current-current correlation function of the Hubbard model with $U/t=8$ at half-filling 
vs real time at three different temperatures. The solid (dashed) lines are results 
after the Wynn (Euler) resummation~(\textit{56}). \textbf{(B)} The imaginary part of the same correlation functions as in (A). 
At long times, where the NLCE convergence is poor, results from 
the two resummations no longer agree and we truncate the correlators before Fourier transforming them to the real-frequency
space. \textbf{(C)} The AC spin conductivity vs real frequency obtained from Eq.~(\ref{eq:sigma3}) (solid lines) and Eq.~(\ref{eq:sigma4}) (dashed lines) 
at the same temperatures as in the first two panels. We take the $\omega=0$ values of the former as our estimate for the DC spin 
conductivity.} 
\label{fig:current2}
\end{figure*}

\subsubsection*{Convergence and the Time Cutoff}

In Fig.~\ref{fig:current2} we show the spin current correlation functions for $U/t=8$ and at three different temperatures as a function of real time, 
and their corresponding conductivities. Although static properties of the model are converged in the NLCE at these temperatures, we find that the 
convergence of dynamical current correlators is lost beyond some time $\tau$ of the order of $1/t$. We identify a small range in time, typically 
$0.1/t$-$0.2/t$ across, in which results from two different resummations of the series begin to deviate from each other. We then choose a time in that range
(see below) as the cutoff time, $\tau_{cutoff}$, before applying 
Eqs.~(\ref{eq:sigma3}) and (\ref{eq:sigma4}). We have checked that possible differences in the resummations near the cutoff time 
lead to negligible differences in the DC spin conductivity. However, the method is unable to capture possible 
contributions to the DC spin conductivity from correlations at times longer than $\tau_{cutoff}$, e.g., at times of the order of $1/J_{ex}$, which become 
especially important at low temperatures.

\begin{figure}[t]
\centering
\includegraphics[width=3.3 in]{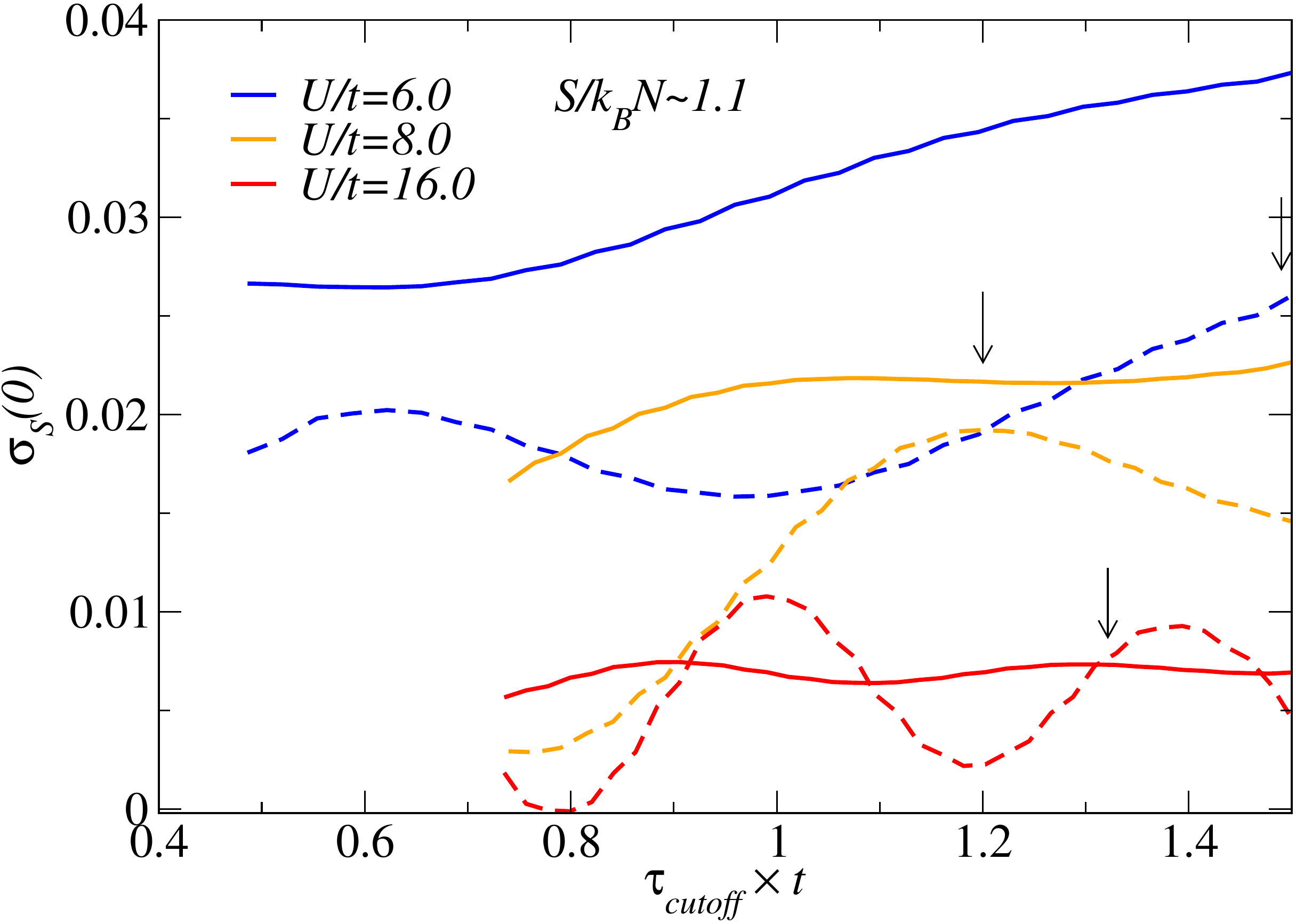}
\caption{{\bf DC spin conductivity as a function of cutoff time.} The DC limit of the spin conductivity 
for $U/t=6$, 8, and 16 obtained via Eq.~(\ref{eq:sigma3}) (solid lines) and Eq.~(\ref{eq:sigma4}) (dashed lines) as a
function of the cutoff time used for the current correlators. The results are for an entropy per particle of about 1.1. The vertical 
arrows denote the location of the cutoff time used for the final estimate based on a set of criteria (see text).}
\label{fig:cut}
\end{figure}

To demonstrate, Fig.~\ref{fig:cut} 
shows the evolution of the DC spin conductivity as $\tau_{cutoff}$ varies over a wide range that captures one or two 
periods of oscillation of the current-current correlation function within or near the NLCE convergence region 
for three different interaction strengths. Vertical arrows denote the final cutoff time chosen for each $U/t$ based on the criteria described below.
Solid and dashed lines are results from Eqs.~(\ref{eq:sigma3}) and (\ref{eq:sigma4}), respectively. 
They point to the insufficient cutoff time and large discrepancy between results of our two equations for the conductivity in the 
weak-coupling region ($t/U\gtrsim 0.125$). Note that the dependence of the DC conductivity on the cutoff time cannot be reliably predicted at longer times
as other time scales, such as $1/J_{ex}$, may affect the functionality in nontrivial ways. Therefore, we treat our estimate for the DC spin
conductivity based on the finite cutoff times as lower bounds.

Equations~(\ref{eq:sigma3}) and (\ref{eq:sigma4}) can hint at the uncertainty in $\sigma_S(\omega)$ 
due to the truncation of the long time tail of the current correlators. The $\omega=0$ limit of $\sigma_S$ from Eq.~(\ref{eq:sigma3})
contains only the real part of the current correlator whereas Eq.~(\ref{eq:sigma4})
contains only the imaginary part and is much more sensitively dependent on its long time tail. 
The real and imaginary parts of the correlator are related through the 
Kramers-Kronig relation and we should expect the same $\sigma_S$ using either formula. However, the time
cutoff can result in a disagreement between $\sigma_S$ obtained 
from the two equations, especially in the DC limit. We choose the cutoff time in the range described above such that
this disagreement is minimized. This generally coincides with the minimum errors in the $f$-sum rule for the spin conductivity (see below). 
Nevertheless, we find that the disagreement between the DC spin conductivity obtained from Eqs.~(\ref{eq:sigma3}) and (\ref{eq:sigma4}) 
can be as large as 30\% for an entropy per particle of $\sim 1.1$ and is largest in the weak-coupling region.

\subsubsection*{Other Checks and Sum Rules}

Even though it is not particularly sensitive to the value of conductivity in the DC limit, we use the $f$-sum rule for 
spins, expressed as~(\textit{60}) 
\begin{equation}
\int_0^{\infty}\textrm{Re\,}\sigma_S(\omega)d\omega = -\frac{\pi\bra{k_x}\ket}{8},
\end{equation}
where $\bra{k_x}\ket$ is the kinetic energy along the $x$-direction, to check for the accuracy of our AC conductivity. 
We find that the relative error is always less than 3\%.

As another check of the accuracy of $\sigma_S(\omega)$, we take advantage of the following equation
\begin{equation}
\Lambda_S(\mathcal{T}) = \int_{-\infty}^{\infty} \frac{d\omega}{\pi} \frac{-e^{-\mathcal{T}\omega}}{1-e^{-\beta\omega}}  \textrm{ Im }\Lambda_S(\omega),
\end{equation}
where $-2\textrm{Im } \Lambda_S(\omega)=2\omega \textrm{Re } \sigma_S(\omega)$ is the spectral function of the retarded spin 
current-current correlation function, $\Lambda_S(\mathcal{T}) = - \left<T_{\mathcal{T}} J_S(\mathcal{T})J_S(0)\right>$ is the 
retarded Green's function of the spin current operator in the imaginary time space, which can be calculated exactly in the NLCE, 
$\mathcal{T}$ is the imaginary time, and
$T_{\mathcal{T}}$ is the imaginary time ordering operator. At $\mathcal{T}=\beta/2$,  
we can simplify the integral
\begin{equation}
\Lambda_S(\beta/2) = \int_{-\infty}^{\infty} \frac{d\omega}{\pi}\  \frac{\omega}{2\sinh (\beta\omega/2)}  \textrm{ Re }\sigma_S(\omega), \label{eq:lambdas} 
\end{equation}
which at low enough temperatures can be approximated to be $\frac{\pi}{\beta^2}\sigma_S(0)$~(\textit{61}). 
We calculate the left hand side directly within the 
NLCE and independently calculate the right hand side using the $\textrm{Re } \sigma_S(\omega)$ we obtained from Eqs.
~(\ref{eq:sigma3}) and (\ref{eq:sigma4}). We find that $\textrm{Re } \sigma_S(\omega)$ from Eq.~(\ref{eq:sigma3}) leads to a $\Lambda_S(\beta/2)$ 
that is consistently within a few percent of the exact value, while the one from Eq.~(\ref{eq:sigma4}) leads to significantly larger errors. 
Hence, we choose the values obtained through Eq.~(\ref{eq:sigma3}) to represent $\sigma_S(0)$.
The spin diffusivity is in turn calculated using the Einstein relation and the exact uniform spin susceptibility we  
calculate using the NLCE.

\begin{figure}[t]
\centering
\includegraphics[width=3.3 in]{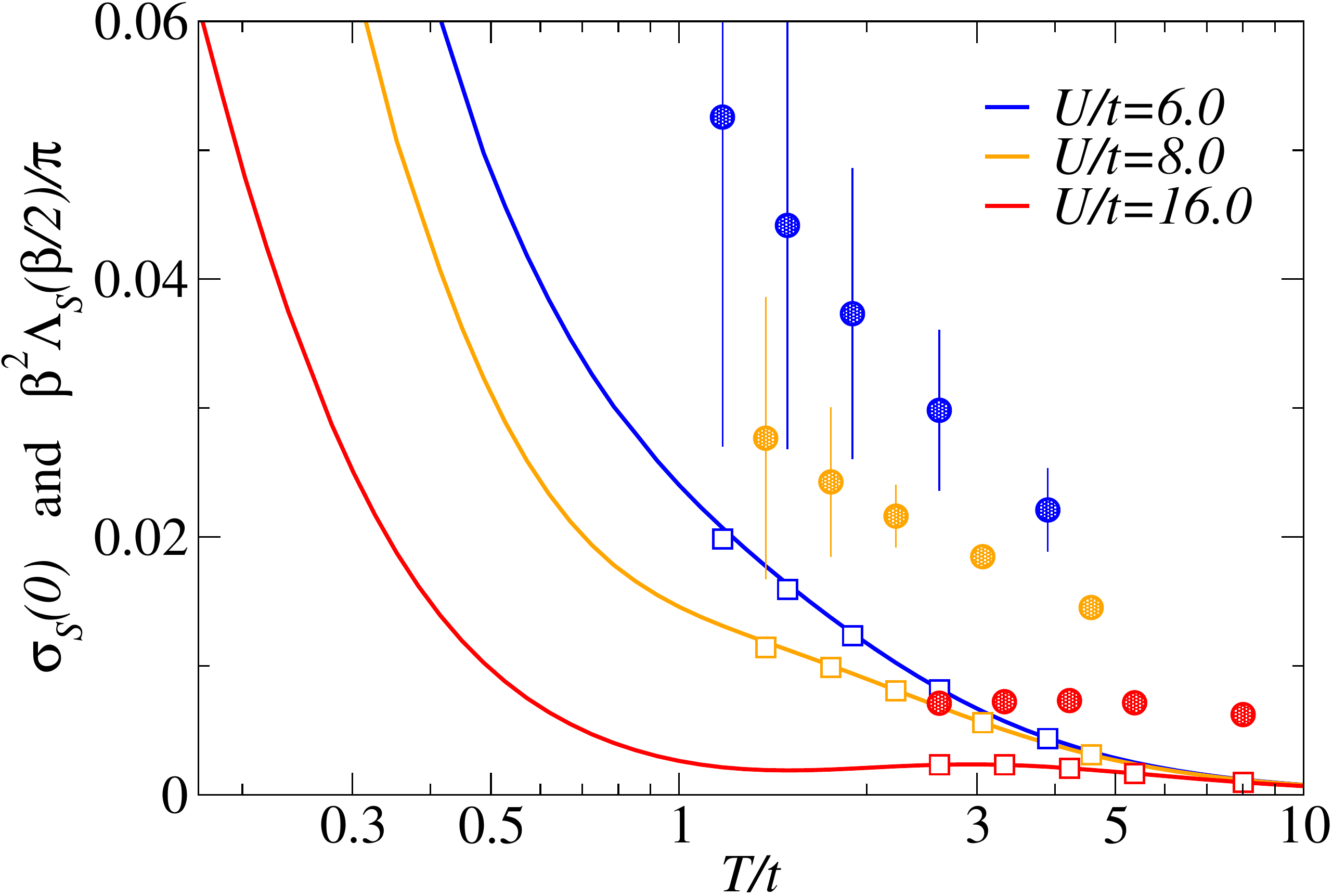}
\caption{{\bf Spin conductivity vs temperature.} Circles are spin conductivities of the Hubbard model at half-filling for three values 
of the interaction strength calculated using Eq.~(\ref{eq:sigma3}). Error bars represent the difference between the latter and the corresponding estimates
obtained using Eq.~(\ref{eq:sigma4}). Lines of the same color show the corresponding $\frac{\beta^2}{\pi}\Lambda_S(\beta/2)$, 
calculated directly in the NLCE. Squares are the right hand side of Eq.~(\ref{eq:lambdas}), calculated using the AC spin 
conductivity obtained from Eq.~(\ref{eq:sigma3}), multiplied by $\frac{\beta^2}{\pi}$.}
\label{fig:cond}
\end{figure}

In Fig.~\ref{fig:cond} we show the DC spin conductivity as a function of temperature for $U/t=6$, 8 and 16. We also plot $\frac{\beta^2}{\pi}\Lambda_S(\beta/2)$ and the right hand side of Eq.~(\ref{eq:lambdas}) multiplied by $\frac{\beta^2}{\pi}$ as a function of temperature as solid lines and squares, respectively. It is clear that $\frac{\beta^2}{\pi}\Lambda_S(\beta/2)$ can serve only as a proxy here, and cannot replace a direct calculation of $\sigma_S(0)$.


\begin{thebibliography}{61}%
\makeatletter
\providecommand \@ifxundefined [1]{%
 \@ifx{#1\undefined}
}%
\providecommand \@ifnum [1]{%
 \ifnum #1\expandafter \@firstoftwo
 \else \expandafter \@secondoftwo
 \fi
}%
\providecommand \@ifx [1]{%
 \ifx #1\expandafter \@firstoftwo
 \else \expandafter \@secondoftwo
 \fi
}%
\providecommand \natexlab [1]{#1}%
\providecommand \enquote  [1]{``#1''}%
\providecommand \bibnamefont  [1]{#1}%
\providecommand \bibfnamefont [1]{#1}%
\providecommand \citenamefont [1]{#1}%
\providecommand \href@noop [0]{\@secondoftwo}%
\providecommand \href [0]{\begingroup \@sanitize@url \@href}%
\providecommand \@href[1]{\@@startlink{#1}\@@href}%
\providecommand \@@href[1]{\endgroup#1\@@endlink}%
\providecommand \@sanitize@url [0]{\catcode `\\12\catcode `\$12\catcode
  `\&12\catcode `\#12\catcode `\^12\catcode `\_12\catcode `\%12\relax}%
\providecommand \@@startlink[1]{}%
\providecommand \@@endlink[0]{}%
\providecommand \url  [0]{\begingroup\@sanitize@url \@url }%
\providecommand \@url [1]{\endgroup\@href {#1}{\urlprefix }}%
\providecommand \urlprefix  [0]{URL }%
\providecommand \Eprint [0]{\href }%
\providecommand \doibase [0]{http://dx.doi.org/}%
\providecommand \selectlanguage [0]{\@gobble}%
\providecommand \bibinfo  [0]{\@secondoftwo}%
\providecommand \bibfield  [0]{\@secondoftwo}%
\providecommand \translation [1]{[#1]}%
\providecommand \BibitemOpen [0]{}%
\providecommand \bibitemStop [0]{}%
\providecommand \bibitemNoStop [0]{.\EOS\space}%
\providecommand \EOS [0]{\spacefactor3000\relax}%
\providecommand \BibitemShut  [1]{\csname bibitem#1\endcsname}%
\let\auto@bib@innerbib\@empty
\bibitem [{\citenamefont {Giamarchi}(2004)}]{Giamarchi2004}%
  \BibitemOpen
  \bibfield  {author} {\bibinfo {author} {\bibfnamefont {T.}~\bibnamefont
  {Giamarchi}},\ }\href@noop {} {\emph {\bibinfo {title} {Quantum Physics in
  One Dimension}}}\ (\bibinfo  {publisher} {Clarendon Press},\ \bibinfo {year}
  {2004})\BibitemShut {NoStop}%
\bibitem [{\citenamefont {Auslaender}\ \emph {et~al.}(2005)\citenamefont
  {Auslaender}, \citenamefont {Steinberg}, \citenamefont {Yacoby},
  \citenamefont {Tserkovnyak}, \citenamefont {Halperin}, \citenamefont
  {Baldwin}, \citenamefont {Pfeiffer},\ and\ \citenamefont
  {West}}]{Auslaender2005}%
  \BibitemOpen
  \bibfield  {author} {\bibinfo {author} {\bibfnamefont {O.~M.}\ \bibnamefont
  {Auslaender}}, \bibinfo {author} {\bibfnamefont {H.}~\bibnamefont
  {Steinberg}}, \bibinfo {author} {\bibfnamefont {A.}~\bibnamefont {Yacoby}},
  \bibinfo {author} {\bibfnamefont {Y.}~\bibnamefont {Tserkovnyak}}, \bibinfo
  {author} {\bibfnamefont {B.~I.}\ \bibnamefont {Halperin}}, \bibinfo {author}
  {\bibfnamefont {K.~W.}\ \bibnamefont {Baldwin}}, \bibinfo {author}
  {\bibfnamefont {L.~N.}\ \bibnamefont {Pfeiffer}}, \ and\ \bibinfo {author}
  {\bibfnamefont {K.~W.}\ \bibnamefont {West}},\ }\href {\doibase
  10.1126/science.1107821} {\bibfield  {journal} {\bibinfo  {journal}
  {Science}\ }\textbf {\bibinfo {volume} {308}},\ \bibinfo {pages} {88}
  (\bibinfo {year} {2005})}\BibitemShut {NoStop}%
\bibitem [{\citenamefont {Heeger}\ \emph {et~al.}(1988)\citenamefont {Heeger},
  \citenamefont {Kivelson}, \citenamefont {Schrieffer},\ and\ \citenamefont
  {Su}}]{Heeger1988SSH}%
  \BibitemOpen
  \bibfield  {author} {\bibinfo {author} {\bibfnamefont {A.~J.}\ \bibnamefont
  {Heeger}}, \bibinfo {author} {\bibfnamefont {S.}~\bibnamefont {Kivelson}},
  \bibinfo {author} {\bibfnamefont {J.~R.}\ \bibnamefont {Schrieffer}}, \ and\
  \bibinfo {author} {\bibfnamefont {W.~P.}\ \bibnamefont {Su}},\ }\href@noop {}
  {\bibfield  {journal} {\bibinfo  {journal} {Rev. Mod. Phys.}\ }\textbf
  {\bibinfo {volume} {60}},\ \bibinfo {pages} {781} (\bibinfo {year}
  {1988})}\BibitemShut {NoStop}%
\bibitem [{\citenamefont {Anderson}(1997)}]{Anderson1997}%
  \BibitemOpen
  \bibfield  {author} {\bibinfo {author} {\bibfnamefont {P.~W.}\ \bibnamefont
  {Anderson}},\ }\href@noop {} {\bibfield  {journal} {\bibinfo  {journal}
  {Physics Today}\ }\textbf {\bibinfo {volume} {50}},\ \bibinfo {pages} {42}
  (\bibinfo {year} {1997})}\BibitemShut {NoStop}%
\bibitem [{\citenamefont {Lee}\ \emph {et~al.}(2006)\citenamefont {Lee},
  \citenamefont {Nagaosa},\ and\ \citenamefont {Wen}}]{lee06hightc}%
  \BibitemOpen
  \bibfield  {author} {\bibinfo {author} {\bibfnamefont {P.~A.}\ \bibnamefont
  {Lee}}, \bibinfo {author} {\bibfnamefont {N.}~\bibnamefont {Nagaosa}}, \ and\
  \bibinfo {author} {\bibfnamefont {X.-G.}\ \bibnamefont {Wen}},\ }\href@noop
  {} {\bibfield  {journal} {\bibinfo  {journal} {Rev. Mod. Phys.}\ }\textbf
  {\bibinfo {volume} {78}},\ \bibinfo {pages} {17} (\bibinfo {year}
  {2006})}\BibitemShut {NoStop}%
\bibitem [{\citenamefont {Lieb}\ and\ \citenamefont {Wu}(1968)}]{Lieb1968}%
  \BibitemOpen
  \bibfield  {author} {\bibinfo {author} {\bibfnamefont {E.~H.}\ \bibnamefont
  {Lieb}}\ and\ \bibinfo {author} {\bibfnamefont {F.~Y.}\ \bibnamefont {Wu}},\
  }\href {\doibase 10.1103/PhysRevLett.20.1445} {\bibfield  {journal} {\bibinfo
   {journal} {Phys. Rev. Lett.}\ }\textbf {\bibinfo {volume} {20}},\ \bibinfo
  {pages} {1445} (\bibinfo {year} {1968})}\BibitemShut {NoStop}%
\bibitem [{\citenamefont {Scalapino}\ \emph {et~al.}(1993)\citenamefont
  {Scalapino}, \citenamefont {White},\ and\ \citenamefont
  {Zhang}}]{Scalapino1993}%
  \BibitemOpen
  \bibfield  {author} {\bibinfo {author} {\bibfnamefont {D.~J.}\ \bibnamefont
  {Scalapino}}, \bibinfo {author} {\bibfnamefont {S.~R.}\ \bibnamefont
  {White}}, \ and\ \bibinfo {author} {\bibfnamefont {S.}~\bibnamefont
  {Zhang}},\ }\href {\doibase 10.1103/PhysRevB.47.7995} {\bibfield  {journal}
  {\bibinfo  {journal} {Phys. Rev. B}\ }\textbf {\bibinfo {volume} {47}},\
  \bibinfo {pages} {7995} (\bibinfo {year} {1993})}\BibitemShut {NoStop}%
\bibitem [{\citenamefont {Bon\ifmmode~\check{c}\else \v{c}\fi{}a}\ and\
  \citenamefont {Jakli\ifmmode~\check{c}\else \v{c}\fi{}}(1995)}]{Bonca1995}%
  \BibitemOpen
  \bibfield  {author} {\bibinfo {author} {\bibfnamefont {J.}~\bibnamefont
  {Bon\ifmmode~\check{c}\else \v{c}\fi{}a}}\ and\ \bibinfo {author}
  {\bibfnamefont {J.}~\bibnamefont {Jakli\ifmmode~\check{c}\else \v{c}\fi{}}},\
  }\href {\doibase 10.1103/PhysRevB.51.16083} {\bibfield  {journal} {\bibinfo
  {journal} {Phys. Rev. B}\ }\textbf {\bibinfo {volume} {51}},\ \bibinfo
  {pages} {16083} (\bibinfo {year} {1995})}\BibitemShut {NoStop}%
\bibitem [{\citenamefont {Kopietz}(1998)}]{Kopietz1998}%
  \BibitemOpen
  \bibfield  {author} {\bibinfo {author} {\bibfnamefont {P.}~\bibnamefont
  {Kopietz}},\ }\href {\doibase 10.1103/PhysRevB.57.7829} {\bibfield  {journal}
  {\bibinfo  {journal} {Phys. Rev. B}\ }\textbf {\bibinfo {volume} {57}},\
  \bibinfo {pages} {7829} (\bibinfo {year} {1998})}\BibitemShut {NoStop}%
\bibitem [{\citenamefont {Mukerjee}\ \emph {et~al.}(2006)\citenamefont
  {Mukerjee}, \citenamefont {Oganesyan},\ and\ \citenamefont
  {Huse}}]{Mukerjee2006}%
  \BibitemOpen
  \bibfield  {author} {\bibinfo {author} {\bibfnamefont {S.}~\bibnamefont
  {Mukerjee}}, \bibinfo {author} {\bibfnamefont {V.}~\bibnamefont {Oganesyan}},
  \ and\ \bibinfo {author} {\bibfnamefont {D.}~\bibnamefont {Huse}},\ }\href
  {\doibase 10.1103/PhysRevB.73.035113} {\bibfield  {journal} {\bibinfo
  {journal} {Phys. Rev. B}\ }\textbf {\bibinfo {volume} {73}},\ \bibinfo
  {pages} {035113} (\bibinfo {year} {2006})}\BibitemShut {NoStop}%
\bibitem [{\citenamefont {Kim}\ and\ \citenamefont
  {Huse}(2012)}]{Hyungwon2012}%
  \BibitemOpen
  \bibfield  {author} {\bibinfo {author} {\bibfnamefont {H.}~\bibnamefont
  {Kim}}\ and\ \bibinfo {author} {\bibfnamefont {D.~A.}\ \bibnamefont {Huse}},\
  }\href {\doibase 10.1103/PhysRevA.86.053607} {\bibfield  {journal} {\bibinfo
  {journal} {Phys. Rev. A}\ }\textbf {\bibinfo {volume} {86}},\ \bibinfo
  {pages} {053607} (\bibinfo {year} {2012})}\BibitemShut {NoStop}%
\bibitem [{\citenamefont {Snyder}\ and\ \citenamefont
  {De~Silva}(2012)}]{Snyder2012}%
  \BibitemOpen
  \bibfield  {author} {\bibinfo {author} {\bibfnamefont {A.~P.}\ \bibnamefont
  {Snyder}}\ and\ \bibinfo {author} {\bibfnamefont {T.~N.}\ \bibnamefont
  {De~Silva}},\ }\href {\doibase 10.1103/PhysRevA.86.053610} {\bibfield
  {journal} {\bibinfo  {journal} {Phys. Rev. A}\ }\textbf {\bibinfo {volume}
  {86}},\ \bibinfo {pages} {053610} (\bibinfo {year} {2012})}\BibitemShut
  {NoStop}%
\bibitem [{\citenamefont {Karrasch}\ \emph {et~al.}(2014)\citenamefont
  {Karrasch}, \citenamefont {Kennes},\ and\ \citenamefont
  {Moore}}]{Karrasch2014}%
  \BibitemOpen
  \bibfield  {author} {\bibinfo {author} {\bibfnamefont {C.}~\bibnamefont
  {Karrasch}}, \bibinfo {author} {\bibfnamefont {D.~M.}\ \bibnamefont
  {Kennes}}, \ and\ \bibinfo {author} {\bibfnamefont {J.~E.}\ \bibnamefont
  {Moore}},\ }\href {\doibase 10.1103/PhysRevB.90.155104} {\bibfield  {journal}
  {\bibinfo  {journal} {Phys. Rev. B}\ }\textbf {\bibinfo {volume} {90}},\
  \bibinfo {pages} {155104} (\bibinfo {year} {2014})}\BibitemShut {NoStop}%
\bibitem [{\citenamefont {Esslinger}(2010)}]{Esslinger2010FermiHubbard}%
  \BibitemOpen
  \bibfield  {author} {\bibinfo {author} {\bibfnamefont {T.}~\bibnamefont
  {Esslinger}},\ }\href@noop {} {\bibfield  {journal} {\bibinfo  {journal}
  {Annu. Rev. Condens. Matter Phys.}\ }\textbf {\bibinfo {volume} {1}},\
  \bibinfo {pages} {129} (\bibinfo {year} {2010})}\BibitemShut {NoStop}%
\bibitem [{\citenamefont {Cheuk}\ \emph {et~al.}(2015)\citenamefont {Cheuk},
  \citenamefont {Nichols}, \citenamefont {Okan}, \citenamefont {Gersdorf},
  \citenamefont {Ramasesh}, \citenamefont {Bakr}, \citenamefont {Lompe},\ and\
  \citenamefont {Zwierlein}}]{Cheuk2015}%
  \BibitemOpen
  \bibfield  {author} {\bibinfo {author} {\bibfnamefont {L.~W.}\ \bibnamefont
  {Cheuk}}, \bibinfo {author} {\bibfnamefont {M.~A.}\ \bibnamefont {Nichols}},
  \bibinfo {author} {\bibfnamefont {M.}~\bibnamefont {Okan}}, \bibinfo {author}
  {\bibfnamefont {T.}~\bibnamefont {Gersdorf}}, \bibinfo {author}
  {\bibfnamefont {V.~V.}\ \bibnamefont {Ramasesh}}, \bibinfo {author}
  {\bibfnamefont {W.~S.}\ \bibnamefont {Bakr}}, \bibinfo {author}
  {\bibfnamefont {T.}~\bibnamefont {Lompe}}, \ and\ \bibinfo {author}
  {\bibfnamefont {M.~W.}\ \bibnamefont {Zwierlein}},\ }\href@noop {} {\bibfield
   {journal} {\bibinfo  {journal} {Phys. Rev. Lett.}\ }\textbf {\bibinfo
  {volume} {114}},\ \bibinfo {pages} {193001} (\bibinfo {year}
  {2015})}\BibitemShut {NoStop}%
\bibitem [{\citenamefont {Haller}\ \emph {et~al.}(2015)\citenamefont {Haller},
  \citenamefont {Hudson}, \citenamefont {Kelly}, \citenamefont {Cotta},
  \citenamefont {Peaudecerf}, \citenamefont {Bruce},\ and\ \citenamefont
  {Kuhr}}]{Haller2015}%
  \BibitemOpen
  \bibfield  {author} {\bibinfo {author} {\bibfnamefont {E.}~\bibnamefont
  {Haller}}, \bibinfo {author} {\bibfnamefont {J.}~\bibnamefont {Hudson}},
  \bibinfo {author} {\bibfnamefont {A.}~\bibnamefont {Kelly}}, \bibinfo
  {author} {\bibfnamefont {D.~A.}\ \bibnamefont {Cotta}}, \bibinfo {author}
  {\bibfnamefont {B.}~\bibnamefont {Peaudecerf}}, \bibinfo {author}
  {\bibfnamefont {G.~D.}\ \bibnamefont {Bruce}}, \ and\ \bibinfo {author}
  {\bibfnamefont {S.}~\bibnamefont {Kuhr}},\ }\href@noop {} {\bibfield
  {journal} {\bibinfo  {journal} {Nat. Phys.}\ }\textbf {\bibinfo {volume}
  {11}},\ \bibinfo {pages} {738} (\bibinfo {year} {2015})}\BibitemShut
  {NoStop}%
\bibitem [{\citenamefont {Parsons}\ \emph {et~al.}(2015)\citenamefont
  {Parsons}, \citenamefont {Huber}, \citenamefont {Mazurenko}, \citenamefont
  {Chiu}, \citenamefont {Setiawan}, \citenamefont {Wooley-Brown}, \citenamefont
  {Blatt},\ and\ \citenamefont {Greiner}}]{Parsons2015}%
  \BibitemOpen
  \bibfield  {author} {\bibinfo {author} {\bibfnamefont {M.~F.}\ \bibnamefont
  {Parsons}}, \bibinfo {author} {\bibfnamefont {F.}~\bibnamefont {Huber}},
  \bibinfo {author} {\bibfnamefont {A.}~\bibnamefont {Mazurenko}}, \bibinfo
  {author} {\bibfnamefont {C.~S.}\ \bibnamefont {Chiu}}, \bibinfo {author}
  {\bibfnamefont {W.}~\bibnamefont {Setiawan}}, \bibinfo {author}
  {\bibfnamefont {K.}~\bibnamefont {Wooley-Brown}}, \bibinfo {author}
  {\bibfnamefont {S.}~\bibnamefont {Blatt}}, \ and\ \bibinfo {author}
  {\bibfnamefont {M.}~\bibnamefont {Greiner}},\ }\href@noop {} {\bibfield
  {journal} {\bibinfo  {journal} {Phys. Rev. Lett.}\ }\textbf {\bibinfo
  {volume} {114}},\ \bibinfo {pages} {213002} (\bibinfo {year}
  {2015})}\BibitemShut {NoStop}%
\bibitem [{\citenamefont {Omran}\ \emph {et~al.}(2015)\citenamefont {Omran},
  \citenamefont {Boll}, \citenamefont {Hilker}, \citenamefont {Kleinlein},
  \citenamefont {Salomon}, \citenamefont {Bloch},\ and\ \citenamefont
  {Gross}}]{Omran2015}%
  \BibitemOpen
  \bibfield  {author} {\bibinfo {author} {\bibfnamefont {A.}~\bibnamefont
  {Omran}}, \bibinfo {author} {\bibfnamefont {M.}~\bibnamefont {Boll}},
  \bibinfo {author} {\bibfnamefont {T.~A.}\ \bibnamefont {Hilker}}, \bibinfo
  {author} {\bibfnamefont {K.}~\bibnamefont {Kleinlein}}, \bibinfo {author}
  {\bibfnamefont {G.}~\bibnamefont {Salomon}}, \bibinfo {author} {\bibfnamefont
  {I.}~\bibnamefont {Bloch}}, \ and\ \bibinfo {author} {\bibfnamefont
  {C.}~\bibnamefont {Gross}},\ }\href@noop {} {\bibfield  {journal} {\bibinfo
  {journal} {Phys. Rev. Lett.}\ }\textbf {\bibinfo {volume} {115}},\ \bibinfo
  {pages} {263001} (\bibinfo {year} {2015})}\BibitemShut {NoStop}%
\bibitem [{\citenamefont {Edge}\ \emph {et~al.}(2015)\citenamefont {Edge},
  \citenamefont {Anderson}, \citenamefont {Jervis}, \citenamefont {McKay},
  \citenamefont {Day}, \citenamefont {Trotzky},\ and\ \citenamefont
  {Thywissen}}]{Edge2015}%
  \BibitemOpen
  \bibfield  {author} {\bibinfo {author} {\bibfnamefont {G.~J.~A.}\
  \bibnamefont {Edge}}, \bibinfo {author} {\bibfnamefont {R.}~\bibnamefont
  {Anderson}}, \bibinfo {author} {\bibfnamefont {D.}~\bibnamefont {Jervis}},
  \bibinfo {author} {\bibfnamefont {D.~C.}\ \bibnamefont {McKay}}, \bibinfo
  {author} {\bibfnamefont {R.}~\bibnamefont {Day}}, \bibinfo {author}
  {\bibfnamefont {S.}~\bibnamefont {Trotzky}}, \ and\ \bibinfo {author}
  {\bibfnamefont {J.~H.}\ \bibnamefont {Thywissen}},\ }\href@noop {} {\bibfield
   {journal} {\bibinfo  {journal} {Phys. Rev. A}\ }\textbf {\bibinfo {volume}
  {92}},\ \bibinfo {pages} {063406} (\bibinfo {year} {2015})}\BibitemShut
  {NoStop}%
\bibitem [{\citenamefont {Brown}\ \emph {et~al.}(2017)\citenamefont {Brown},
  \citenamefont {Mitra}, \citenamefont {Guardado-Sanchez}, \citenamefont
  {Schau{\ss}}, \citenamefont {Kondov}, \citenamefont {Khatami}, \citenamefont
  {Paiva}, \citenamefont {Trivedi}, \citenamefont {Huse},\ and\ \citenamefont
  {Bakr}}]{Brown2017}%
  \BibitemOpen
  \bibfield  {author} {\bibinfo {author} {\bibfnamefont {P.~T.}\ \bibnamefont
  {Brown}}, \bibinfo {author} {\bibfnamefont {D.}~\bibnamefont {Mitra}},
  \bibinfo {author} {\bibfnamefont {E.}~\bibnamefont {Guardado-Sanchez}},
  \bibinfo {author} {\bibfnamefont {P.}~\bibnamefont {Schau{\ss}}}, \bibinfo
  {author} {\bibfnamefont {S.~S.}\ \bibnamefont {Kondov}}, \bibinfo {author}
  {\bibfnamefont {E.}~\bibnamefont {Khatami}}, \bibinfo {author} {\bibfnamefont
  {T.}~\bibnamefont {Paiva}}, \bibinfo {author} {\bibfnamefont
  {N.}~\bibnamefont {Trivedi}}, \bibinfo {author} {\bibfnamefont {D.~A.}\
  \bibnamefont {Huse}}, \ and\ \bibinfo {author} {\bibfnamefont {W.~S.}\
  \bibnamefont {Bakr}},\ }\href {\doibase 10.1126/science.aam7838} {\bibfield
  {journal} {\bibinfo  {journal} {Science}\ }\textbf {\bibinfo {volume}
  {357}},\ \bibinfo {pages} {1385} (\bibinfo {year} {2017})}\BibitemShut
  {NoStop}%
\bibitem [{\citenamefont {Cocchi}\ \emph {et~al.}(2016)\citenamefont {Cocchi},
  \citenamefont {Miller}, \citenamefont {Drewes}, \citenamefont {Koschorreck},
  \citenamefont {Pertot}, \citenamefont {Brennecke},\ and\ \citenamefont
  {K\"ohl}}]{Cocchi2015}%
  \BibitemOpen
  \bibfield  {author} {\bibinfo {author} {\bibfnamefont {E.}~\bibnamefont
  {Cocchi}}, \bibinfo {author} {\bibfnamefont {L.~A.}\ \bibnamefont {Miller}},
  \bibinfo {author} {\bibfnamefont {J.~H.}\ \bibnamefont {Drewes}}, \bibinfo
  {author} {\bibfnamefont {M.}~\bibnamefont {Koschorreck}}, \bibinfo {author}
  {\bibfnamefont {D.}~\bibnamefont {Pertot}}, \bibinfo {author} {\bibfnamefont
  {F.}~\bibnamefont {Brennecke}}, \ and\ \bibinfo {author} {\bibfnamefont
  {M.}~\bibnamefont {K\"ohl}},\ }\href@noop {} {\bibfield  {journal} {\bibinfo
  {journal} {Phys. Rev. Lett.}\ }\textbf {\bibinfo {volume} {116}},\ \bibinfo
  {pages} {175301} (\bibinfo {year} {2016})}\BibitemShut {NoStop}%
\bibitem [{\citenamefont {Hofrichter}\ \emph {et~al.}(2016)\citenamefont
  {Hofrichter}, \citenamefont {Riegger}, \citenamefont {Scazza}, \citenamefont
  {H\"ofer}, \citenamefont {Fernandes}, \citenamefont {Bloch},\ and\
  \citenamefont {F\"olling}}]{hofrichter2015}%
  \BibitemOpen
  \bibfield  {author} {\bibinfo {author} {\bibfnamefont {C.}~\bibnamefont
  {Hofrichter}}, \bibinfo {author} {\bibfnamefont {L.}~\bibnamefont {Riegger}},
  \bibinfo {author} {\bibfnamefont {F.}~\bibnamefont {Scazza}}, \bibinfo
  {author} {\bibfnamefont {M.}~\bibnamefont {H\"ofer}}, \bibinfo {author}
  {\bibfnamefont {D.~R.}\ \bibnamefont {Fernandes}}, \bibinfo {author}
  {\bibfnamefont {I.}~\bibnamefont {Bloch}}, \ and\ \bibinfo {author}
  {\bibfnamefont {S.}~\bibnamefont {F\"olling}},\ }\href@noop {} {\bibfield
  {journal} {\bibinfo  {journal} {Phys. Rev. X}\ }\textbf {\bibinfo {volume}
  {6}},\ \bibinfo {pages} {021030} (\bibinfo {year} {2016})}\BibitemShut
  {NoStop}%
\bibitem [{\citenamefont {Cheuk}\ \emph
  {et~al.}(2016{\natexlab{a}})\citenamefont {Cheuk}, \citenamefont {Nichols},
  \citenamefont {Lawrence}, \citenamefont {Okan}, \citenamefont {Zhang},
  \citenamefont {Khatami}, \citenamefont {Trivedi}, \citenamefont {Paiva},
  \citenamefont {Rigol},\ and\ \citenamefont
  {Zwierlein}}]{CheukSpinCharge2016}%
  \BibitemOpen
  \bibfield  {author} {\bibinfo {author} {\bibfnamefont {L.~W.}\ \bibnamefont
  {Cheuk}}, \bibinfo {author} {\bibfnamefont {M.~A.}\ \bibnamefont {Nichols}},
  \bibinfo {author} {\bibfnamefont {K.~R.}\ \bibnamefont {Lawrence}}, \bibinfo
  {author} {\bibfnamefont {M.}~\bibnamefont {Okan}}, \bibinfo {author}
  {\bibfnamefont {H.}~\bibnamefont {Zhang}}, \bibinfo {author} {\bibfnamefont
  {E.}~\bibnamefont {Khatami}}, \bibinfo {author} {\bibfnamefont
  {N.}~\bibnamefont {Trivedi}}, \bibinfo {author} {\bibfnamefont
  {T.}~\bibnamefont {Paiva}}, \bibinfo {author} {\bibfnamefont
  {M.}~\bibnamefont {Rigol}}, \ and\ \bibinfo {author} {\bibfnamefont {M.~W.}\
  \bibnamefont {Zwierlein}},\ }\href {\doibase 10.1126/science.aag3349}
  {\bibfield  {journal} {\bibinfo  {journal} {Science}\ }\textbf {\bibinfo
  {volume} {353}},\ \bibinfo {pages} {1260} (\bibinfo {year}
  {2016}{\natexlab{a}})}\BibitemShut {NoStop}%
\bibitem [{\citenamefont {Boll}\ \emph {et~al.}(2016)\citenamefont {Boll},
  \citenamefont {Hilker}, \citenamefont {Salomon}, \citenamefont {Omran},
  \citenamefont {Nespolo}, \citenamefont {Pollet}, \citenamefont {Bloch},\ and\
  \citenamefont {Gross}}]{Boll2016}%
  \BibitemOpen
  \bibfield  {author} {\bibinfo {author} {\bibfnamefont {M.}~\bibnamefont
  {Boll}}, \bibinfo {author} {\bibfnamefont {T.~A.}\ \bibnamefont {Hilker}},
  \bibinfo {author} {\bibfnamefont {G.}~\bibnamefont {Salomon}}, \bibinfo
  {author} {\bibfnamefont {A.}~\bibnamefont {Omran}}, \bibinfo {author}
  {\bibfnamefont {J.}~\bibnamefont {Nespolo}}, \bibinfo {author} {\bibfnamefont
  {L.}~\bibnamefont {Pollet}}, \bibinfo {author} {\bibfnamefont
  {I.}~\bibnamefont {Bloch}}, \ and\ \bibinfo {author} {\bibfnamefont
  {C.}~\bibnamefont {Gross}},\ }\href {\doibase 10.1126/science.aag1635}
  {\bibfield  {journal} {\bibinfo  {journal} {Science}\ }\textbf {\bibinfo
  {volume} {353}},\ \bibinfo {pages} {1257} (\bibinfo {year}
  {2016})}\BibitemShut {NoStop}%
\bibitem [{\citenamefont {Parsons}\ \emph {et~al.}(2016)\citenamefont
  {Parsons}, \citenamefont {Mazurenko}, \citenamefont {Chiu}, \citenamefont
  {Ji}, \citenamefont {Greif},\ and\ \citenamefont {Greiner}}]{Parsons2016}%
  \BibitemOpen
  \bibfield  {author} {\bibinfo {author} {\bibfnamefont {M.~F.}\ \bibnamefont
  {Parsons}}, \bibinfo {author} {\bibfnamefont {A.}~\bibnamefont {Mazurenko}},
  \bibinfo {author} {\bibfnamefont {C.~S.}\ \bibnamefont {Chiu}}, \bibinfo
  {author} {\bibfnamefont {G.}~\bibnamefont {Ji}}, \bibinfo {author}
  {\bibfnamefont {D.}~\bibnamefont {Greif}}, \ and\ \bibinfo {author}
  {\bibfnamefont {M.}~\bibnamefont {Greiner}},\ }\href {\doibase
  10.1126/science.aag1430} {\bibfield  {journal} {\bibinfo  {journal}
  {Science}\ }\textbf {\bibinfo {volume} {353}},\ \bibinfo {pages} {1253}
  (\bibinfo {year} {2016})}\BibitemShut {NoStop}%
\bibitem [{\citenamefont {Cheneau}\ \emph {et~al.}(2012)\citenamefont
  {Cheneau}, \citenamefont {Barmettler}, \citenamefont {Poletti}, \citenamefont
  {Endres}, \citenamefont {Schau\ss}, \citenamefont {Fukuhara}, \citenamefont
  {Gross}, \citenamefont {Bloch}, \citenamefont {Kollath},\ and\ \citenamefont
  {Kuhr}}]{Cheneau2012Lightcone}%
  \BibitemOpen
  \bibfield  {author} {\bibinfo {author} {\bibfnamefont {M.}~\bibnamefont
  {Cheneau}}, \bibinfo {author} {\bibfnamefont {P.}~\bibnamefont {Barmettler}},
  \bibinfo {author} {\bibfnamefont {D.}~\bibnamefont {Poletti}}, \bibinfo
  {author} {\bibfnamefont {M.}~\bibnamefont {Endres}}, \bibinfo {author}
  {\bibfnamefont {P.}~\bibnamefont {Schau\ss}}, \bibinfo {author}
  {\bibfnamefont {T.}~\bibnamefont {Fukuhara}}, \bibinfo {author}
  {\bibfnamefont {C.}~\bibnamefont {Gross}}, \bibinfo {author} {\bibfnamefont
  {I.}~\bibnamefont {Bloch}}, \bibinfo {author} {\bibfnamefont
  {C.}~\bibnamefont {Kollath}}, \ and\ \bibinfo {author} {\bibfnamefont
  {S.}~\bibnamefont {Kuhr}},\ }\href@noop {} {\bibfield  {journal} {\bibinfo
  {journal} {Nature}\ }\textbf {\bibinfo {volume} {481}},\ \bibinfo {pages}
  {484} (\bibinfo {year} {2012})}\BibitemShut {NoStop}%
\bibitem [{\citenamefont {Fukuhara}\ \emph
  {et~al.}(2013{\natexlab{a}})\citenamefont {Fukuhara}, \citenamefont
  {Kantian}, \citenamefont {Endres}, \citenamefont {Cheneau}, \citenamefont
  {Schau{\ss}}, \citenamefont {Hild}, \citenamefont {Bellem}, \citenamefont
  {Schollw{\"o}ck}, \citenamefont {Giamarchi}, \citenamefont {Gross},
  \citenamefont {Bloch},\ and\ \citenamefont {Kuhr}}]{FukuharaImpurity2013}%
  \BibitemOpen
  \bibfield  {author} {\bibinfo {author} {\bibfnamefont {T.}~\bibnamefont
  {Fukuhara}}, \bibinfo {author} {\bibfnamefont {A.}~\bibnamefont {Kantian}},
  \bibinfo {author} {\bibfnamefont {M.}~\bibnamefont {Endres}}, \bibinfo
  {author} {\bibfnamefont {M.}~\bibnamefont {Cheneau}}, \bibinfo {author}
  {\bibfnamefont {P.}~\bibnamefont {Schau{\ss}}}, \bibinfo {author}
  {\bibfnamefont {S.}~\bibnamefont {Hild}}, \bibinfo {author} {\bibfnamefont
  {D.}~\bibnamefont {Bellem}}, \bibinfo {author} {\bibfnamefont
  {U.}~\bibnamefont {Schollw{\"o}ck}}, \bibinfo {author} {\bibfnamefont
  {T.}~\bibnamefont {Giamarchi}}, \bibinfo {author} {\bibfnamefont
  {C.}~\bibnamefont {Gross}}, \bibinfo {author} {\bibfnamefont
  {I.}~\bibnamefont {Bloch}}, \ and\ \bibinfo {author} {\bibfnamefont
  {S.}~\bibnamefont {Kuhr}},\ }\href@noop {} {\bibfield  {journal} {\bibinfo
  {journal} {Nature Physics}\ }\textbf {\bibinfo {volume} {9}},\ \bibinfo
  {pages} {235} (\bibinfo {year} {2013}{\natexlab{a}})}\BibitemShut {NoStop}%
\bibitem [{\citenamefont {Fukuhara}\ \emph
  {et~al.}(2013{\natexlab{b}})\citenamefont {Fukuhara}, \citenamefont
  {Schau\ss{}}, \citenamefont {Endres}, \citenamefont {Hild}, \citenamefont
  {Cheneau}, \citenamefont {Bloch},\ and\ \citenamefont
  {Gross}}]{Fukuhara2013}%
  \BibitemOpen
  \bibfield  {author} {\bibinfo {author} {\bibfnamefont {T.}~\bibnamefont
  {Fukuhara}}, \bibinfo {author} {\bibfnamefont {P.}~\bibnamefont
  {Schau\ss{}}}, \bibinfo {author} {\bibfnamefont {M.}~\bibnamefont {Endres}},
  \bibinfo {author} {\bibfnamefont {S.}~\bibnamefont {Hild}}, \bibinfo {author}
  {\bibfnamefont {M.}~\bibnamefont {Cheneau}}, \bibinfo {author} {\bibfnamefont
  {I.}~\bibnamefont {Bloch}}, \ and\ \bibinfo {author} {\bibfnamefont
  {C.}~\bibnamefont {Gross}},\ }\href@noop {} {\bibfield  {journal} {\bibinfo
  {journal} {Nature}\ }\textbf {\bibinfo {volume} {502}},\ \bibinfo {pages}
  {76} (\bibinfo {year} {2013}{\natexlab{b}})}\BibitemShut {NoStop}%
\bibitem [{\citenamefont {Hild}\ \emph {et~al.}(2014)\citenamefont {Hild},
  \citenamefont {Fukuhara}, \citenamefont {Schau\ss{}}, \citenamefont {Zeiher},
  \citenamefont {Knap}, \citenamefont {Demler}, \citenamefont {Bloch},\ and\
  \citenamefont {Gross}}]{Hild2014}%
  \BibitemOpen
  \bibfield  {author} {\bibinfo {author} {\bibfnamefont {S.}~\bibnamefont
  {Hild}}, \bibinfo {author} {\bibfnamefont {T.}~\bibnamefont {Fukuhara}},
  \bibinfo {author} {\bibfnamefont {P.}~\bibnamefont {Schau\ss{}}}, \bibinfo
  {author} {\bibfnamefont {J.}~\bibnamefont {Zeiher}}, \bibinfo {author}
  {\bibfnamefont {M.}~\bibnamefont {Knap}}, \bibinfo {author} {\bibfnamefont
  {E.}~\bibnamefont {Demler}}, \bibinfo {author} {\bibfnamefont
  {I.}~\bibnamefont {Bloch}}, \ and\ \bibinfo {author} {\bibfnamefont
  {C.}~\bibnamefont {Gross}},\ }\href {\doibase 10.1103/PhysRevLett.113.147205}
  {\bibfield  {journal} {\bibinfo  {journal} {Phys. Rev. Lett.}\ }\textbf
  {\bibinfo {volume} {113}},\ \bibinfo {pages} {147205} (\bibinfo {year}
  {2014})}\BibitemShut {NoStop}%
\bibitem [{\citenamefont {Preiss}\ \emph {et~al.}(2015)\citenamefont {Preiss},
  \citenamefont {Ma}, \citenamefont {Tai}, \citenamefont {Lukin}, \citenamefont
  {Rispoli}, \citenamefont {Zupancic}, \citenamefont {Lahini}, \citenamefont
  {Islam},\ and\ \citenamefont {Greiner}}]{Preiss2015}%
  \BibitemOpen
  \bibfield  {author} {\bibinfo {author} {\bibfnamefont {P.~M.}\ \bibnamefont
  {Preiss}}, \bibinfo {author} {\bibfnamefont {R.}~\bibnamefont {Ma}}, \bibinfo
  {author} {\bibfnamefont {M.~E.}\ \bibnamefont {Tai}}, \bibinfo {author}
  {\bibfnamefont {A.}~\bibnamefont {Lukin}}, \bibinfo {author} {\bibfnamefont
  {M.}~\bibnamefont {Rispoli}}, \bibinfo {author} {\bibfnamefont
  {P.}~\bibnamefont {Zupancic}}, \bibinfo {author} {\bibfnamefont
  {Y.}~\bibnamefont {Lahini}}, \bibinfo {author} {\bibfnamefont
  {R.}~\bibnamefont {Islam}}, \ and\ \bibinfo {author} {\bibfnamefont
  {M.}~\bibnamefont {Greiner}},\ }\href {\doibase 10.1126/science.1260364}
  {\bibfield  {journal} {\bibinfo  {journal} {Science}\ }\textbf {\bibinfo
  {volume} {347}},\ \bibinfo {pages} {1229} (\bibinfo {year}
  {2015})}\BibitemShut {NoStop}%
\bibitem [{\citenamefont {Choi}\ \emph {et~al.}(2016)\citenamefont {Choi},
  \citenamefont {Hild}, \citenamefont {Zeiher}, \citenamefont {Schau{\ss}},
  \citenamefont {Rubio-Abadal}, \citenamefont {Yefsah}, \citenamefont
  {Khemani}, \citenamefont {Huse}, \citenamefont {Bloch},\ and\ \citenamefont
  {Gross}}]{Choi2016}%
  \BibitemOpen
  \bibfield  {author} {\bibinfo {author} {\bibfnamefont {J.-y.}\ \bibnamefont
  {Choi}}, \bibinfo {author} {\bibfnamefont {S.}~\bibnamefont {Hild}}, \bibinfo
  {author} {\bibfnamefont {J.}~\bibnamefont {Zeiher}}, \bibinfo {author}
  {\bibfnamefont {P.}~\bibnamefont {Schau{\ss}}}, \bibinfo {author}
  {\bibfnamefont {A.}~\bibnamefont {Rubio-Abadal}}, \bibinfo {author}
  {\bibfnamefont {T.}~\bibnamefont {Yefsah}}, \bibinfo {author} {\bibfnamefont
  {V.}~\bibnamefont {Khemani}}, \bibinfo {author} {\bibfnamefont {D.~A.}\
  \bibnamefont {Huse}}, \bibinfo {author} {\bibfnamefont {I.}~\bibnamefont
  {Bloch}}, \ and\ \bibinfo {author} {\bibfnamefont {C.}~\bibnamefont
  {Gross}},\ }\href {\doibase 10.1126/science.aaf8834} {\bibfield  {journal}
  {\bibinfo  {journal} {Science}\ }\textbf {\bibinfo {volume} {352}},\ \bibinfo
  {pages} {1547} (\bibinfo {year} {2016})}\BibitemShut {NoStop}%
\bibitem [{\citenamefont {Strohmaier}\ \emph {et~al.}(2007)\citenamefont
  {Strohmaier}, \citenamefont {Takasu}, \citenamefont {G{\"u}nter},
  \citenamefont {J{\"o}rdens}, \citenamefont {K{\"o}hl}, \citenamefont
  {Moritz},\ and\ \citenamefont {Esslinger}}]{stro07transport}%
  \BibitemOpen
  \bibfield  {author} {\bibinfo {author} {\bibfnamefont {N.}~\bibnamefont
  {Strohmaier}}, \bibinfo {author} {\bibfnamefont {Y.}~\bibnamefont {Takasu}},
  \bibinfo {author} {\bibfnamefont {K.}~\bibnamefont {G{\"u}nter}}, \bibinfo
  {author} {\bibfnamefont {R.}~\bibnamefont {J{\"o}rdens}}, \bibinfo {author}
  {\bibfnamefont {M.}~\bibnamefont {K{\"o}hl}}, \bibinfo {author}
  {\bibfnamefont {H.}~\bibnamefont {Moritz}}, \ and\ \bibinfo {author}
  {\bibfnamefont {T.}~\bibnamefont {Esslinger}},\ }\href@noop {} {\bibfield
  {journal} {\bibinfo  {journal} {Phys. Rev. Lett.}\ }\textbf {\bibinfo
  {volume} {99}},\ \bibinfo {pages} {220601} (\bibinfo {year}
  {2007})}\BibitemShut {NoStop}%
\bibitem [{\citenamefont {Schneider}\ \emph {et~al.}(2012)\citenamefont
  {Schneider}, \citenamefont {Hackerm\"uller}, \citenamefont {Ronzheimer},
  \citenamefont {Will}, \citenamefont {Braun}, \citenamefont {Best},
  \citenamefont {Bloch}, \citenamefont {Demler}, \citenamefont {Mandt},
  \citenamefont {Rasch},\ and\ \citenamefont {Rosch}}]{Schneider2012}%
  \BibitemOpen
  \bibfield  {author} {\bibinfo {author} {\bibfnamefont {U.}~\bibnamefont
  {Schneider}}, \bibinfo {author} {\bibfnamefont {L.}~\bibnamefont
  {Hackerm\"uller}}, \bibinfo {author} {\bibfnamefont {J.~P.}\ \bibnamefont
  {Ronzheimer}}, \bibinfo {author} {\bibfnamefont {S.}~\bibnamefont {Will}},
  \bibinfo {author} {\bibfnamefont {S.}~\bibnamefont {Braun}}, \bibinfo
  {author} {\bibfnamefont {T.}~\bibnamefont {Best}}, \bibinfo {author}
  {\bibfnamefont {I.}~\bibnamefont {Bloch}}, \bibinfo {author} {\bibfnamefont
  {E.}~\bibnamefont {Demler}}, \bibinfo {author} {\bibfnamefont
  {S.}~\bibnamefont {Mandt}}, \bibinfo {author} {\bibfnamefont
  {D.}~\bibnamefont {Rasch}}, \ and\ \bibinfo {author} {\bibfnamefont
  {A.}~\bibnamefont {Rosch}},\ }\href@noop {} {\bibfield  {journal} {\bibinfo
  {journal} {Nat. Phys.}\ }\textbf {\bibinfo {volume} {8}},\ \bibinfo {pages}
  {213} (\bibinfo {year} {2012})}\BibitemShut {NoStop}%
\bibitem [{\citenamefont {Xu}\ \emph {et~al.}(2018)\citenamefont {Xu},
  \citenamefont {McGehee}, \citenamefont {Morong},\ and\ \citenamefont
  {DeMarco}}]{WXu2016}%
  \BibitemOpen
  \bibfield  {author} {\bibinfo {author} {\bibfnamefont {W.}~\bibnamefont
  {Xu}}, \bibinfo {author} {\bibfnamefont {W.~R.}\ \bibnamefont {McGehee}},
  \bibinfo {author} {\bibfnamefont {W.~N.}\ \bibnamefont {Morong}}, \ and\
  \bibinfo {author} {\bibfnamefont {B.}~\bibnamefont {DeMarco}},\ }\href@noop
  {} {\bibfield  {journal} {\bibinfo  {journal} {ArXiv e-prints}\ } (\bibinfo
  {year} {2018})},\ \Eprint {http://arxiv.org/abs/1606.06669v5}
  {arXiv:1606.06669v5} \BibitemShut {NoStop}%
\bibitem [{\citenamefont {Anderson}\ \emph {et~al.}(2018)\citenamefont
  {Anderson}, \citenamefont {Wang}, \citenamefont {Xu}, \citenamefont {Venu},
  \citenamefont {Trotzky}, \citenamefont {Chevy},\ and\ \citenamefont
  {Thywissen}}]{RAnderson2017}%
  \BibitemOpen
  \bibfield  {author} {\bibinfo {author} {\bibfnamefont {R.}~\bibnamefont
  {Anderson}}, \bibinfo {author} {\bibfnamefont {F.}~\bibnamefont {Wang}},
  \bibinfo {author} {\bibfnamefont {P.}~\bibnamefont {Xu}}, \bibinfo {author}
  {\bibfnamefont {V.}~\bibnamefont {Venu}}, \bibinfo {author} {\bibfnamefont
  {S.}~\bibnamefont {Trotzky}}, \bibinfo {author} {\bibfnamefont
  {F.}~\bibnamefont {Chevy}}, \ and\ \bibinfo {author} {\bibfnamefont {J.~H.}\
  \bibnamefont {Thywissen}},\ }\href@noop {} {\bibfield  {journal} {\bibinfo
  {journal} {ArXiv e-prints}\ } (\bibinfo {year} {2018})},\ \Eprint
  {http://arxiv.org/abs/1712.09965v2} {arXiv:1712.09965v2} \BibitemShut
  {NoStop}%
\bibitem [{\citenamefont {Lebrat}\ \emph {et~al.}(2018)\citenamefont {Lebrat},
  \citenamefont {Gri\ifmmode~\check{s}\else \v{s}\fi{}ins}, \citenamefont
  {Husmann}, \citenamefont {H\"ausler}, \citenamefont {Corman}, \citenamefont
  {Giamarchi}, \citenamefont {Brantut},\ and\ \citenamefont
  {Esslinger}}]{Lebrat2017}%
  \BibitemOpen
  \bibfield  {author} {\bibinfo {author} {\bibfnamefont {M.}~\bibnamefont
  {Lebrat}}, \bibinfo {author} {\bibfnamefont {P.}~\bibnamefont
  {Gri\ifmmode~\check{s}\else \v{s}\fi{}ins}}, \bibinfo {author} {\bibfnamefont
  {D.}~\bibnamefont {Husmann}}, \bibinfo {author} {\bibfnamefont
  {S.}~\bibnamefont {H\"ausler}}, \bibinfo {author} {\bibfnamefont
  {L.}~\bibnamefont {Corman}}, \bibinfo {author} {\bibfnamefont
  {T.}~\bibnamefont {Giamarchi}}, \bibinfo {author} {\bibfnamefont {J.-P.}\
  \bibnamefont {Brantut}}, \ and\ \bibinfo {author} {\bibfnamefont
  {T.}~\bibnamefont {Esslinger}},\ }\href {\doibase 10.1103/PhysRevX.8.011053}
  {\bibfield  {journal} {\bibinfo  {journal} {Phys. Rev. X}\ }\textbf {\bibinfo
  {volume} {8}},\ \bibinfo {pages} {011053} (\bibinfo {year}
  {2018})}\BibitemShut {NoStop}%
\bibitem [{\citenamefont {Sommer}\ \emph
  {et~al.}(2011{\natexlab{a}})\citenamefont {Sommer}, \citenamefont {Ku},
  \citenamefont {Roati},\ and\ \citenamefont {Zwierlein}}]{somm11spin}%
  \BibitemOpen
  \bibfield  {author} {\bibinfo {author} {\bibfnamefont {A.}~\bibnamefont
  {Sommer}}, \bibinfo {author} {\bibfnamefont {M.}~\bibnamefont {Ku}}, \bibinfo
  {author} {\bibfnamefont {G.}~\bibnamefont {Roati}}, \ and\ \bibinfo {author}
  {\bibfnamefont {M.~W.}\ \bibnamefont {Zwierlein}},\ }\href@noop {} {\bibfield
   {journal} {\bibinfo  {journal} {Nature}\ }\textbf {\bibinfo {volume}
  {472}},\ \bibinfo {pages} {201} (\bibinfo {year}
  {2011}{\natexlab{a}})}\BibitemShut {NoStop}%
\bibitem [{\citenamefont {Sommer}\ \emph
  {et~al.}(2011{\natexlab{b}})\citenamefont {Sommer}, \citenamefont {Ku},\ and\
  \citenamefont {Zwierlein}}]{somm11polarontransport}%
  \BibitemOpen
  \bibfield  {author} {\bibinfo {author} {\bibfnamefont {A.}~\bibnamefont
  {Sommer}}, \bibinfo {author} {\bibfnamefont {M.}~\bibnamefont {Ku}}, \ and\
  \bibinfo {author} {\bibfnamefont {M.~W.}\ \bibnamefont {Zwierlein}},\
  }\href@noop {} {\bibfield  {journal} {\bibinfo  {journal} {New J. Phys.}\
  }\textbf {\bibinfo {volume} {13}},\ \bibinfo {pages} {055009} (\bibinfo
  {year} {2011}{\natexlab{b}})}\BibitemShut {NoStop}%
\bibitem [{\citenamefont {Bardon}\ \emph {et~al.}(2014)\citenamefont {Bardon},
  \citenamefont {Beattie}, \citenamefont {Luciuk}, \citenamefont {Cairncross},
  \citenamefont {Fine}, \citenamefont {Cheng}, \citenamefont {Edge},
  \citenamefont {Taylor}, \citenamefont {Zhang}, \citenamefont {Trotzky},\ and\
  \citenamefont {Thywissen}}]{bard14}%
  \BibitemOpen
  \bibfield  {author} {\bibinfo {author} {\bibfnamefont {A.~B.}\ \bibnamefont
  {Bardon}}, \bibinfo {author} {\bibfnamefont {S.}~\bibnamefont {Beattie}},
  \bibinfo {author} {\bibfnamefont {C.}~\bibnamefont {Luciuk}}, \bibinfo
  {author} {\bibfnamefont {W.}~\bibnamefont {Cairncross}}, \bibinfo {author}
  {\bibfnamefont {D.}~\bibnamefont {Fine}}, \bibinfo {author} {\bibfnamefont
  {N.~S.}\ \bibnamefont {Cheng}}, \bibinfo {author} {\bibfnamefont {G.~J.~A.}\
  \bibnamefont {Edge}}, \bibinfo {author} {\bibfnamefont {E.}~\bibnamefont
  {Taylor}}, \bibinfo {author} {\bibfnamefont {S.}~\bibnamefont {Zhang}},
  \bibinfo {author} {\bibfnamefont {S.}~\bibnamefont {Trotzky}}, \ and\
  \bibinfo {author} {\bibfnamefont {J.~H.}\ \bibnamefont {Thywissen}},\
  }\href@noop {} {\bibfield  {journal} {\bibinfo  {journal} {Science}\ }\textbf
  {\bibinfo {volume} {344}},\ \bibinfo {pages} {722} (\bibinfo {year}
  {2014})}\BibitemShut {NoStop}%
\bibitem [{\citenamefont {Valtolina}\ \emph {et~al.}(2017)\citenamefont
  {Valtolina}, \citenamefont {Scazza}, \citenamefont {Amico}, \citenamefont
  {Burchianti}, \citenamefont {Recati}, \citenamefont {Enss}, \citenamefont
  {Inguscio}, \citenamefont {Zaccanti},\ and\ \citenamefont
  {Roati}}]{valtolina2017}%
  \BibitemOpen
  \bibfield  {author} {\bibinfo {author} {\bibfnamefont {G.}~\bibnamefont
  {Valtolina}}, \bibinfo {author} {\bibfnamefont {F.}~\bibnamefont {Scazza}},
  \bibinfo {author} {\bibfnamefont {A.}~\bibnamefont {Amico}}, \bibinfo
  {author} {\bibfnamefont {A.}~\bibnamefont {Burchianti}}, \bibinfo {author}
  {\bibfnamefont {A.}~\bibnamefont {Recati}}, \bibinfo {author} {\bibfnamefont
  {T.}~\bibnamefont {Enss}}, \bibinfo {author} {\bibfnamefont {M.}~\bibnamefont
  {Inguscio}}, \bibinfo {author} {\bibfnamefont {M.}~\bibnamefont {Zaccanti}},
  \ and\ \bibinfo {author} {\bibfnamefont {G.}~\bibnamefont {Roati}},\
  }\href@noop {} {\bibfield  {journal} {\bibinfo  {journal} {Nat. Phys.}\
  }\textbf {\bibinfo {volume} {13}},\ \bibinfo {pages} {704} (\bibinfo {year}
  {2017})}\BibitemShut {NoStop}%
\bibitem [{\citenamefont {Koschorreck}\ \emph {et~al.}(2013)\citenamefont
  {Koschorreck}, \citenamefont {Pertot}, \citenamefont {Vogt},\ and\
  \citenamefont {K{\"o}hl}}]{koschorreck13}%
  \BibitemOpen
  \bibfield  {author} {\bibinfo {author} {\bibfnamefont {M.}~\bibnamefont
  {Koschorreck}}, \bibinfo {author} {\bibfnamefont {D.}~\bibnamefont {Pertot}},
  \bibinfo {author} {\bibfnamefont {E.}~\bibnamefont {Vogt}}, \ and\ \bibinfo
  {author} {\bibfnamefont {M.}~\bibnamefont {K{\"o}hl}},\ }\href@noop {}
  {\bibfield  {journal} {\bibinfo  {journal} {Nat Phys}\ }\textbf {\bibinfo
  {volume} {9}},\ \bibinfo {pages} {405} (\bibinfo {year} {2013})}\BibitemShut
  {NoStop}%
\bibitem [{\citenamefont {Luciuk}\ \emph {et~al.}(2017)\citenamefont {Luciuk},
  \citenamefont {Smale}, \citenamefont {B\"ottcher}, \citenamefont {Sharum},
  \citenamefont {Olsen}, \citenamefont {Trotzky}, \citenamefont {Enss},\ and\
  \citenamefont {Thywissen}}]{Luciuk2017}%
  \BibitemOpen
  \bibfield  {author} {\bibinfo {author} {\bibfnamefont {C.}~\bibnamefont
  {Luciuk}}, \bibinfo {author} {\bibfnamefont {S.}~\bibnamefont {Smale}},
  \bibinfo {author} {\bibfnamefont {F.}~\bibnamefont {B\"ottcher}}, \bibinfo
  {author} {\bibfnamefont {H.}~\bibnamefont {Sharum}}, \bibinfo {author}
  {\bibfnamefont {B.~A.}\ \bibnamefont {Olsen}}, \bibinfo {author}
  {\bibfnamefont {S.}~\bibnamefont {Trotzky}}, \bibinfo {author} {\bibfnamefont
  {T.}~\bibnamefont {Enss}}, \ and\ \bibinfo {author} {\bibfnamefont {J.~H.}\
  \bibnamefont {Thywissen}},\ }\href {\doibase 10.1103/PhysRevLett.118.130405}
  {\bibfield  {journal} {\bibinfo  {journal} {Phys. Rev. Lett.}\ }\textbf
  {\bibinfo {volume} {118}},\ \bibinfo {pages} {130405} (\bibinfo {year}
  {2017})}\BibitemShut {NoStop}%
\bibitem [{\citenamefont {Cheuk}\ \emph
  {et~al.}(2016{\natexlab{b}})\citenamefont {Cheuk}, \citenamefont {Nichols},
  \citenamefont {Lawrence}, \citenamefont {Okan}, \citenamefont {Zhang},\ and\
  \citenamefont {Zwierlein}}]{CheukMott2016}%
  \BibitemOpen
  \bibfield  {author} {\bibinfo {author} {\bibfnamefont {L.~W.}\ \bibnamefont
  {Cheuk}}, \bibinfo {author} {\bibfnamefont {M.~A.}\ \bibnamefont {Nichols}},
  \bibinfo {author} {\bibfnamefont {K.~R.}\ \bibnamefont {Lawrence}}, \bibinfo
  {author} {\bibfnamefont {M.}~\bibnamefont {Okan}}, \bibinfo {author}
  {\bibfnamefont {H.}~\bibnamefont {Zhang}}, \ and\ \bibinfo {author}
  {\bibfnamefont {M.~W.}\ \bibnamefont {Zwierlein}},\ }\href@noop {} {\bibfield
   {journal} {\bibinfo  {journal} {Phys. Rev. Lett.}\ }\textbf {\bibinfo
  {volume} {116}},\ \bibinfo {pages} {235301} (\bibinfo {year}
  {2016}{\natexlab{b}})}\BibitemShut {NoStop}%
\bibitem [{sup()}]{supplementalmaterial}%
  \BibitemOpen
  \href@noop {} {\bibinfo  {journal} {See supplementary materials}\ }\BibitemShut
  {NoStop}%
\bibitem [{\citenamefont {DePue}\ \emph {et~al.}(1999)\citenamefont {DePue},
  \citenamefont {McCormick}, \citenamefont {Winoto}, \citenamefont {Oliver},\
  and\ \citenamefont {Weiss}}]{depu99}%
  \BibitemOpen
\bibfield  {journal} {  }\bibfield  {author} {\bibinfo {author} {\bibfnamefont
  {M.~T.}\ \bibnamefont {DePue}}, \bibinfo {author} {\bibfnamefont
  {C.}~\bibnamefont {McCormick}}, \bibinfo {author} {\bibfnamefont {S.~L.}\
  \bibnamefont {Winoto}}, \bibinfo {author} {\bibfnamefont {S.}~\bibnamefont
  {Oliver}}, \ and\ \bibinfo {author} {\bibfnamefont {D.~S.}\ \bibnamefont
  {Weiss}},\ }\href@noop {} {\bibfield  {journal} {\bibinfo  {journal} {Phys.
  Rev. Lett.}\ }\textbf {\bibinfo {volume} {82}},\ \bibinfo {pages} {2262}
  (\bibinfo {year} {1999})}\BibitemShut {NoStop}%
\bibitem [{\citenamefont {Weld}\ \emph {et~al.}(2009)\citenamefont {Weld},
  \citenamefont {Medley}, \citenamefont {Miyake}, \citenamefont {Hucul},
  \citenamefont {Pritchard},\ and\ \citenamefont {Ketterle}}]{Weld2009}%
  \BibitemOpen
  \bibfield  {author} {\bibinfo {author} {\bibfnamefont {D.~M.}\ \bibnamefont
  {Weld}}, \bibinfo {author} {\bibfnamefont {P.}~\bibnamefont {Medley}},
  \bibinfo {author} {\bibfnamefont {H.}~\bibnamefont {Miyake}}, \bibinfo
  {author} {\bibfnamefont {D.}~\bibnamefont {Hucul}}, \bibinfo {author}
  {\bibfnamefont {D.~E.}\ \bibnamefont {Pritchard}}, \ and\ \bibinfo {author}
  {\bibfnamefont {W.}~\bibnamefont {Ketterle}},\ }\href {\doibase
  10.1103/PhysRevLett.103.245301} {\bibfield  {journal} {\bibinfo  {journal}
  {Phys. Rev. Lett.}\ }\textbf {\bibinfo {volume} {103}},\ \bibinfo {pages}
  {245301} (\bibinfo {year} {2009})}\BibitemShut {NoStop}%
\bibitem [{\citenamefont {Batrouni}\ and\ \citenamefont
  {Scalettar}(2017)}]{Batrouni2017}%
  \BibitemOpen
  \bibfield  {author} {\bibinfo {author} {\bibfnamefont {G.~G.}\ \bibnamefont
  {Batrouni}}\ and\ \bibinfo {author} {\bibfnamefont {R.~T.}\ \bibnamefont
  {Scalettar}},\ }\href {\doibase 10.1103/PhysRevA.96.033632} {\bibfield
  {journal} {\bibinfo  {journal} {Phys. Rev. A}\ }\textbf {\bibinfo {volume}
  {96}},\ \bibinfo {pages} {033632} (\bibinfo {year} {2017})}\BibitemShut
  {NoStop}%
\bibitem [{\citenamefont {Rigol}\ \emph {et~al.}(2006)\citenamefont {Rigol},
  \citenamefont {Bryant},\ and\ \citenamefont {Singh}}]{Rigol2006}%
  \BibitemOpen
  \bibfield  {author} {\bibinfo {author} {\bibfnamefont {M.}~\bibnamefont
  {Rigol}}, \bibinfo {author} {\bibfnamefont {T.}~\bibnamefont {Bryant}}, \
  and\ \bibinfo {author} {\bibfnamefont {R.~R.~P.}\ \bibnamefont {Singh}},\
  }\href@noop {} {\bibfield  {journal} {\bibinfo  {journal} {Phys. Rev. Lett.}\
  }\textbf {\bibinfo {volume} {97}},\ \bibinfo {pages} {187202} (\bibinfo
  {year} {2006})}\BibitemShut {NoStop}%
\bibitem [{\citenamefont {Khatami}\ and\ \citenamefont
  {Rigol}(2011)}]{Khatami2011}%
  \BibitemOpen
  \bibfield  {author} {\bibinfo {author} {\bibfnamefont {E.}~\bibnamefont
  {Khatami}}\ and\ \bibinfo {author} {\bibfnamefont {M.}~\bibnamefont
  {Rigol}},\ }\href@noop {} {\bibfield  {journal} {\bibinfo  {journal} {Phys.
  Rev. A}\ }\textbf {\bibinfo {volume} {84}},\ \bibinfo {pages} {053611}
  (\bibinfo {year} {2011})}\BibitemShut {NoStop}%
\bibitem [{\citenamefont {Khatami}\ and\ \citenamefont
  {Rigol}(2012)}]{Khatami2012}%
  \BibitemOpen
  \bibfield  {author} {\bibinfo {author} {\bibfnamefont {E.}~\bibnamefont
  {Khatami}}\ and\ \bibinfo {author} {\bibfnamefont {M.}~\bibnamefont
  {Rigol}},\ }\href@noop {} {\bibfield  {journal} {\bibinfo  {journal} {Phys.
  Rev. A}\ }\textbf {\bibinfo {volume} {86}},\ \bibinfo {pages} {023633}
  (\bibinfo {year} {2012})}\BibitemShut {NoStop}%
\bibitem [{\citenamefont {Bennett}\ and\ \citenamefont
  {Martin}(1965)}]{Bennett1965}%
  \BibitemOpen
  \bibfield  {author} {\bibinfo {author} {\bibfnamefont {H.~S.}\ \bibnamefont
  {Bennett}}\ and\ \bibinfo {author} {\bibfnamefont {P.~C.}\ \bibnamefont
  {Martin}},\ }\href {\doibase 10.1103/PhysRev.138.A608} {\bibfield  {journal}
  {\bibinfo  {journal} {Phys. Rev.}\ }\textbf {\bibinfo {volume} {138}},\
  \bibinfo {pages} {A608} (\bibinfo {year} {1965})}\BibitemShut {NoStop}%
\bibitem [{\citenamefont {Sokol}\ \emph {et~al.}(1993)\citenamefont {Sokol},
  \citenamefont {Gagliano},\ and\ \citenamefont {Bacci}}]{Sokol1993}%
  \BibitemOpen
  \bibfield  {author} {\bibinfo {author} {\bibfnamefont {A.}~\bibnamefont
  {Sokol}}, \bibinfo {author} {\bibfnamefont {E.}~\bibnamefont {Gagliano}}, \
  and\ \bibinfo {author} {\bibfnamefont {S.}~\bibnamefont {Bacci}},\ }\href
  {\doibase 10.1103/PhysRevB.47.14646} {\bibfield  {journal} {\bibinfo
  {journal} {Phys. Rev. B}\ }\textbf {\bibinfo {volume} {47}},\ \bibinfo
  {pages} {14646} (\bibinfo {year} {1993})}\BibitemShut {NoStop}%
\bibitem [{\citenamefont {Ioffe}\ and\ \citenamefont
  {Regel}(1960)}]{Ioffe1960}%
  \BibitemOpen
  \bibfield  {author} {\bibinfo {author} {\bibfnamefont {A.}~\bibnamefont
  {Ioffe}}\ and\ \bibinfo {author} {\bibfnamefont {A.}~\bibnamefont {Regel}},\
  }\href@noop {} {\bibfield  {journal} {\bibinfo  {journal} {Prog. Semicond.}\
  }\textbf {\bibinfo {volume} {4}},\ \bibinfo {pages} {237} (\bibinfo {year}
  {1960})}\BibitemShut {NoStop}%
\bibitem [{\citenamefont {Mott}(1972)}]{Mott1972}%
  \BibitemOpen
  \bibfield  {author} {\bibinfo {author} {\bibfnamefont {N.~F.}\ \bibnamefont
  {Mott}},\ }\href {\doibase 10.1080/14786437208226973} {\bibfield  {journal}
  {\bibinfo  {journal} {The Philosophical Magazine: A Journal of Theoretical
  Experimental and Applied Physics}\ }\textbf {\bibinfo {volume} {26}},\
  \bibinfo {pages} {1015} (\bibinfo {year} {1972})}\BibitemShut {NoStop}%
\bibitem [{\citenamefont {Nichols}\ \emph {et~al.}(2018)\citenamefont
  {Nichols}, \citenamefont {Cheuk}, \citenamefont {Okan}, \citenamefont
  {Hartke}, \citenamefont {Mendez}, \citenamefont {Senthil}, \citenamefont
  {Khatami}, \citenamefont {Zhang},\ and\ \citenamefont
  {Zwierlein}}]{Nichols2018Dataverse}%
  \BibitemOpen
  \bibfield  {author} {\bibinfo {author} {\bibfnamefont {M.~A.}\ \bibnamefont
  {Nichols}}, \bibinfo {author} {\bibfnamefont {L.~W.}\ \bibnamefont {Cheuk}},
  \bibinfo {author} {\bibfnamefont {M.}~\bibnamefont {Okan}}, \bibinfo {author}
  {\bibfnamefont {T.~R.}\ \bibnamefont {Hartke}}, \bibinfo {author}
  {\bibfnamefont {E.}~\bibnamefont {Mendez}}, \bibinfo {author} {\bibfnamefont
  {T.}~\bibnamefont {Senthil}}, \bibinfo {author} {\bibfnamefont
  {E.}~\bibnamefont {Khatami}}, \bibinfo {author} {\bibfnamefont
  {H.}~\bibnamefont {Zhang}}, \ and\ \bibinfo {author} {\bibfnamefont {M.~W.}\
  \bibnamefont {Zwierlein}},\ }\href {\doibase 10.7910/DVN/0OFNFY} {\ Harvard Dataverse
  (\bibinfo {year} {2018}),\ https://doi.org/10.7910/DVN/0OFNFY}\BibitemShut {NoStop}%
\bibitem [{\citenamefont {Tang}\ \emph {et~al.}(2013)\citenamefont {Tang},
  \citenamefont {Khatami},\ and\ \citenamefont {Rigol}}]{Tang2013}%
  \BibitemOpen
  \bibfield  {author} {\bibinfo {author} {\bibfnamefont {B.}~\bibnamefont
  {Tang}}, \bibinfo {author} {\bibfnamefont {E.}~\bibnamefont {Khatami}}, \
  and\ \bibinfo {author} {\bibfnamefont {M.}~\bibnamefont {Rigol}},\
  }\href@noop {} {\bibfield  {journal} {\bibinfo  {journal} {Computer Physics
  Communications}\ }\textbf {\bibinfo {volume} {184}},\ \bibinfo {pages} {557 }
  (\bibinfo {year} {2013})}\BibitemShut {NoStop}%
\bibitem [{\citenamefont {White}\ \emph {et~al.}(2017)\citenamefont {White},
  \citenamefont {Sundar},\ and\ \citenamefont {Hazzard}}]{IGWhite2017}%
  \BibitemOpen
  \bibfield  {author} {\bibinfo {author} {\bibfnamefont {I.~G.}\ \bibnamefont
  {White}}, \bibinfo {author} {\bibfnamefont {B.}~\bibnamefont {Sundar}}, \
  and\ \bibinfo {author} {\bibfnamefont {K.~R.~A.}\ \bibnamefont {Hazzard}},\
  }\href@noop {} {\bibfield  {journal} {\bibinfo  {journal} {ArXiv e-prints}\ }
  (\bibinfo {year} {2017})},\ \Eprint {http://arxiv.org/abs/1710.07696}
  {arXiv:1710.07696} \BibitemShut {NoStop}%
\bibitem [{\citenamefont {Mallayya}\ and\ \citenamefont
  {Rigol}(2018)}]{Mallayya2018}%
  \BibitemOpen
  \bibfield  {author} {\bibinfo {author} {\bibfnamefont {K.}~\bibnamefont
  {Mallayya}}\ and\ \bibinfo {author} {\bibfnamefont {M.}~\bibnamefont
  {Rigol}},\ }\href {\doibase 10.1103/PhysRevLett.120.070603} {\bibfield
  {journal} {\bibinfo  {journal} {Phys. Rev. Lett.}\ }\textbf {\bibinfo
  {volume} {120}},\ \bibinfo {pages} {070603} (\bibinfo {year}
  {2018})}\BibitemShut {NoStop}%
\bibitem [{\citenamefont {Mahan}(2000)}]{Mahan2000}%
  \BibitemOpen
  \bibfield  {author} {\bibinfo {author} {\bibfnamefont {G.~D.}\ \bibnamefont
  {Mahan}},\ }\href@noop {} {\emph {\bibinfo {title} {Many-Particle
  Physics}}},\ \bibinfo {edition} {3rd}\ ed.\ (\bibinfo  {publisher}
  {Springer},\ \bibinfo {address} {New York},\ \bibinfo {year}
  {2000})\BibitemShut {NoStop}%
\bibitem [{\citenamefont {Fishman}\ and\ \citenamefont
  {Jarrell}(2002)}]{Fishman2002}%
  \BibitemOpen
  \bibfield  {author} {\bibinfo {author} {\bibfnamefont {R.~S.}\ \bibnamefont
  {Fishman}}\ and\ \bibinfo {author} {\bibfnamefont {M.}~\bibnamefont
  {Jarrell}},\ }\href {\doibase 10.1063/1.1456431} {\bibfield  {journal}
  {\bibinfo  {journal} {Journal of Applied Physics}\ }\textbf {\bibinfo
  {volume} {91}},\ \bibinfo {pages} {8120} (\bibinfo {year} {2002})}\BibitemShut {NoStop}%
\bibitem [{\citenamefont {Trivedi}\ \emph {et~al.}(1996)\citenamefont
  {Trivedi}, \citenamefont {Scalettar},\ and\ \citenamefont
  {Randeria}}]{Trivedi1996}%
  \BibitemOpen
  \bibfield  {author} {\bibinfo {author} {\bibfnamefont {N.}~\bibnamefont
  {Trivedi}}, \bibinfo {author} {\bibfnamefont {R.~T.}\ \bibnamefont
  {Scalettar}}, \ and\ \bibinfo {author} {\bibfnamefont {M.}~\bibnamefont
  {Randeria}},\ }\href {\doibase 10.1103/PhysRevB.54.R3756} {\bibfield
  {journal} {\bibinfo  {journal} {Phys. Rev. B}\ }\textbf {\bibinfo {volume}
  {54}},\ \bibinfo {pages} {R3756} (\bibinfo {year} {1996})}\BibitemShut
  {NoStop}%
\end{thebibliography}
\end{document}